\documentclass{aastex63}

\accepted{Astroparticle Physics, DOI:10.1016/j.astropartphys.2021.102577 }

\shorttitle{Geometrical interpretation of S5~0716+714 optical variability properties}
\shortauthors{Butuzova M.S.}
\graphicspath{{./}{figures/}}
\usepackage{amsmath}
\usepackage{amssymb}
\usepackage{multirow}

\newcommand{\shiftright}[2]{\makebox[#1][r]{\makebox[0pt][l]{#2}}}

\begin{document}

\title{A geometrical interpretation for the properties of \\multiband optical variability of the blazar S5~0716+714 \footnote{\textcopyright This manuscript version is made available under the CC-BY-NC-ND 4.0 license http://creativecommons.org/licenses/by-nc-nd/4.0/}}

\correspondingauthor{Marina Butuzova}
\email{mbutuzova@craocrimea.ru}

\author[0000-0001-7307-2193]{Marina S. Butuzova}
\affiliation{Crimean Astrophysical Observatory, Nauchny 298409, Crimea, Russia}

\begin{abstract}

We present the results of multiband observations of the blazar S5~0716+714 intra-night variability performed during 23 nights in the period from 04.2014 through 04.2015. 
The bluer-when-brighter trend is detected in both intra- and inter-night data.
We assume that the jet component crossing the region where the medium becomes transparent to the optical radiation forms almost all optical emission of S5~0716+714.
Deviations of some parts of the component from the general trajectory of the component can cause the Doppler factor of these parts to increase.
Various maximum Doppler factors achieved by these parts of the component and different volumes occupied by them with the concave synchrotron self-absorption spectrum result in both the observed various color index behavior in variability and explain the absence of dependence of the bluer-when-brighter behavior on the object magnitude. We estimated spectral maximum frequency $\nu_\text{m}\approx\left( 0.6-1.7\right)\cdot10^{14}$~Hz from intra- and inter-night data.
Assuming that the size of emitting region is comparable with the gravitational radius of a black hole with a mass of $5\cdot10^8$ solar masses, the magnetic field obtained from synchrotron self-absorption is $B\sim10^2-10^4$~G, which corresponds to the values of other independent estimations.
The obtained values of $\nu_\text{m}$ and $B$ confirm our assumption about the nature of the blazar S5~0716+714 region radiating in the optical range.

\end{abstract}

\keywords{galaxies: active galaxies  --- BL Lac objects, Blazars: individual (S5 0716+714) --- radio jets: relativistic jets}

\section{Introduction } 
\label{sec:intro}

The object S5~0716+714 was classified as a blazar due to its strong variability in the entire range of electromagnetic waves, non-thermal emission spectrum, and high polarization degree.
The redshift of S5~0716+714 has not been determined exactly yet.
Various indirect methods have established limits on the redshift: $z\ge0.3$ \citep{Nilsson08, Bychkova06}, $z>0.52$ \citep{Sbarufatti05}, $0.23<z<0.32$ \citep{Danforth13}.
The absence of spectral lines in the blazar radiation and the unresolved underlying galaxy allow us to suggest that almost all the observed emission is formed in the relativistic jet.

Researches of the correlation between the S5~0716+714 variability at different frequencies show a time lag with decreasing observation frequency \citep{Raiteri03, Fuhrmann08, Rani13, Rani14, LiLuo18}.
Only within the optical range, these time lags have not been reliably detected \citep{WuZhou07, Poon09, Dai15, Agarwal16, ZhangWM18}.
It may be due to the small size of the region from which the observed optical radiation comes.
The presence of time lags between optical and radio variability is interpreted by that the regions from which the radiation of the corresponding wavelength comes are spatially separated.
Hence the region radiating in the optical range is closer to the jet base than the radio emission region.
On the one hand, this can be interpreted in terms of the electron energy losses through radiation, as discussed by \citet{Raiteri03}.
Thus, the electrons moving along the jet lose energy and produce synchrotron radiation at lower and lower frequencies.
On the other hand, the spatial separation of regions radiating at different frequencies is naturally explained by the synchrotron self-absorption in the jet  \citep[see, e.g.,][]{Kudr11, Agarwal17}.
Namely, the magnetic field and number density of the radiating particles decrease with moving outwards from the jet base leading to the fact that the medium becomes optically transparent to radiation with a longer and longer wavelength \citep{BlandfordKonigl79, Konigl81, Lobanov98}.
Under the assumption that the brightest and compact feature observed in VLBI maps, called the VLBI core, is the region where the jet medium becomes transparent to radiation at the observed frequency, a shift in the astrometric position of the VLBI cores observed at different radio frequencies is expected.
The first observations of the VLBI core shift were carried out by \citet{MarcaideShapiro84}.
For several hundred active galactic nuclei, including S5~0716+714, it was showed the observed VLBI core shifts to be acceptably explained by the synchrotron self-absorption effect \cite{Pushkarev12, PushKov12, Hov14}. 
\citet{Kovalev05} concluded that the main part of radio emission from the active galactic nuclei, which was observed by single antennae, comes from the VLBI core.
Then the time delay in variability with decreasing observation frequency is because some perturbation, leading to an increase in the flux density, propagates along the jet successively passing through the regions where the jet medium becomes transparent to radiation of a certain frequency and producing a flare on the light curve when being observed at this frequency.
The described passage of the flare was observed for individual objects \citep[e.g., ][]{PushkarevBKH19} and studied for several dozen sources \citep{Plavin19, Kutkin19}.
To date, there is no convincing evidence for any object that the VLBI core cannot be the region in which the jet medium becomes optically thin to radiation at the observation frequency.
Hence, we can expect that there is a region in the jet where the medium becomes transparent to optical radiation, and almost all the observed radiation comes from this region.

Different observed properties of S5~0716+714 (namely, quasi-periods of long-term variability in optical \citep{Raiteri03, Gupta08, Tang12} and radio \citep{Raiteri03, Liu12, Bychkova15, LiLuo18} ranges, and variability of the position angle of the inner part of the parsec-scale jet \citep{Bach05, Lister13}, the kinematics of the VLBI jet features \citep{Rast11, Rani15} were explained under the assumption that the jet from the region radiated in the optical range to 0.15 mas from the 15~GHz VLBI core has the form of a helix with a linearly increasing radius with distance from the jet base \citep{But18a, But18b}.
Based on this result, in this paper, we investigate the intra-day variability (IDV) of the blazar S5~0716+714 during 2014$-$2015.
Section~\ref{sec:obs} includes information about observations and the detected properties of both IDV and long-term variability for 2014-2015.
In Section~\ref{sec:heljet}, we explain the observed stochastic behavior of the flux variability \citep{Amir06, Bhatta13} on different timescales due to geometrical effects.
We show that when taking into account the synchrotron self-absorption and the resulting differentiation of the generation regions of the radiation with different frequencies, an increase in the Doppler factor leads not only to an increase in the radiation flux but also to the bluer-when-brighter (BWB) chromatism.
The possibility of the proposed variability formation scenario and its applicability to the interpretation of the blazar S5~0716+714 variability properties on different timescales are discussed in Section~\ref{sec:discussion}.
Section~\ref{sec:conclusions} contains the conclusions.

\section{Observations}
\label{sec:obs}

Photometric observations of S5~0716+714 were performed using the Maksutov telescope AZT-5 at the Crimean Observatory of SAI MSU.
The diameters of the spherical mirror and the corrective meniscus are 70 and 50~cm, respectively, and the focal length of the system is 200~cm.
The telescope was equipped with the 3326$\times$2504 pixel CCD-camera Alta~U8300.
The image was binned to 2$\times$2 pixels; this yielded a scale of 1.08$^{\prime\prime}$ per pixel.
The field of view of the CCD-camera is 30$^\prime \times$22$^\prime$.
The CCD-camera was cooled to $-20^\circ$~C.
In front of the camera, there was a block of filters that correspond to the B, V filters in the Johnson system and the R, I filters in the Cousins system.
Every night, before and after observations of the object, a series of bias and dark current images were obtained with the maximum exposure corresponding to the exposure in the B filter.
Once a week we performed a series of flat-field images at the time of the evening and morning twilight sky.
Bias, dark and flat-field images were cleared from emissions of the so-called``hot'' and ``cold'' particles by comparing a large number of files obtained throughout the night.
The stages of processing the acquired images are described in detail by \citet{Doroshenko2005}.
Photometry of the object and comparison stars was performed with an aperture of 15$^{\prime\prime}$.
Stars 8, 10, and 18 in the close vicinity of the blazar were selected as comparison stars whose stellar magnitudes are given in \citep{Doroshenko2005}.
These stars correspond to stars 3, 2, and 6 in \citep{Villata98}.
The total duration of observations is almost 107 hours, the average time resolution -- 266 seconds, and the best of it is 131 seconds.
We take into account Galactic extinction towards S5~0716+714 with the coefficient presented in \citep{SchlaflyFink2011}.

\subsection{Intra-day variability}
\begin{figure}
\begin{minipage}[h]{0.49\linewidth}
\center{\includegraphics[width=1\linewidth]{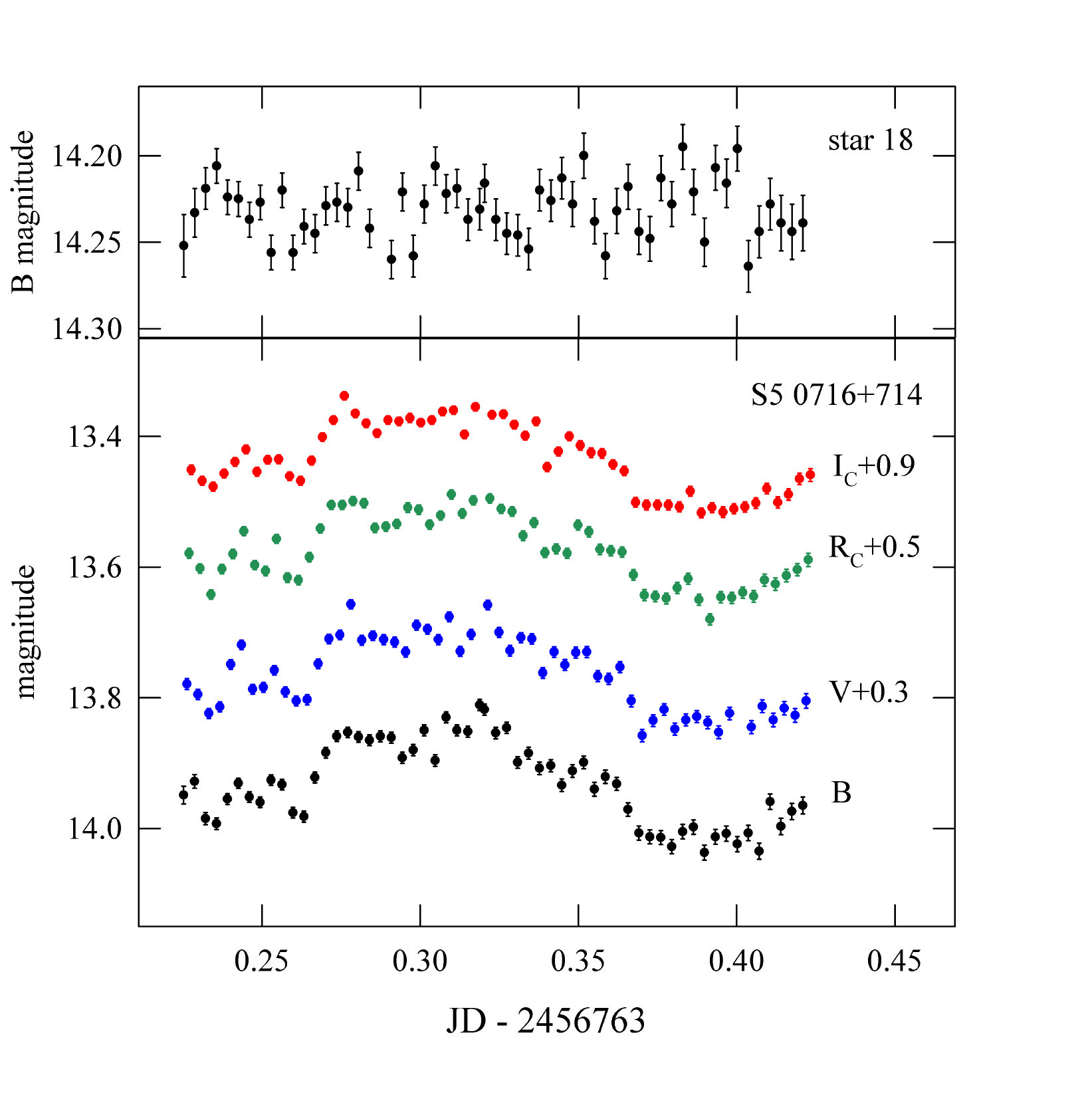}} 
\end{minipage}
\hfill
\begin{minipage}[h]{0.49\linewidth}
\center{\includegraphics[width=1\linewidth]{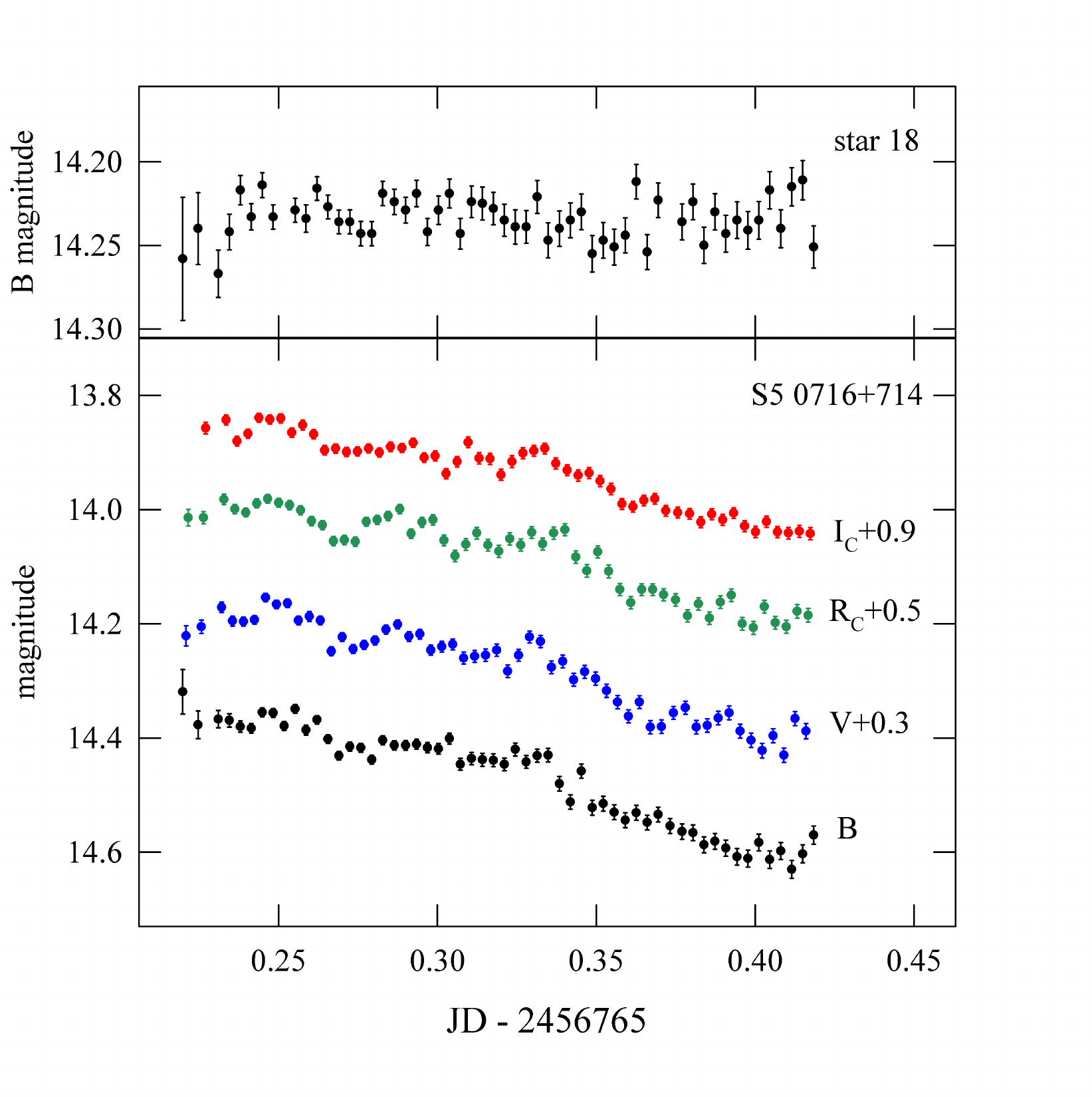}} 
\end{minipage}
\caption{Multiband intra-day light curves of S5~0716+714 for nights with the detected IDV. For better visualization, we added 0.3, 0.5, 0.9 to V, R$_\text{C}$, I$_\text{C}$ magnitudes, respectively. The full figure is available in Appendix.}
\label{fig:lc}
\end{figure}

To detect IDV, we performed the analysis of variance (ANOVA) \citep{deDiego98, deDiego10}.
To do this, we have divided the intra-day data into groups. 
One group included 5 consecutive in time measurements.
If the last group had less than 5 measurements, these were added to the previous group.
ANOVA was applied to both the object and the comparison star to test the brightness stability of the latter and the sky quality.
Table~\ref{tab:IDVtable} shows the results of ANOVA tests.
The intra-night variability is present throughout 16 nights out of 23.
Intra-day light curves for dates with the detected IDV are plotted in Fig.~\ref{fig:lc}.

\begin{figure}
    \centering
    \includegraphics[scale=0.7]{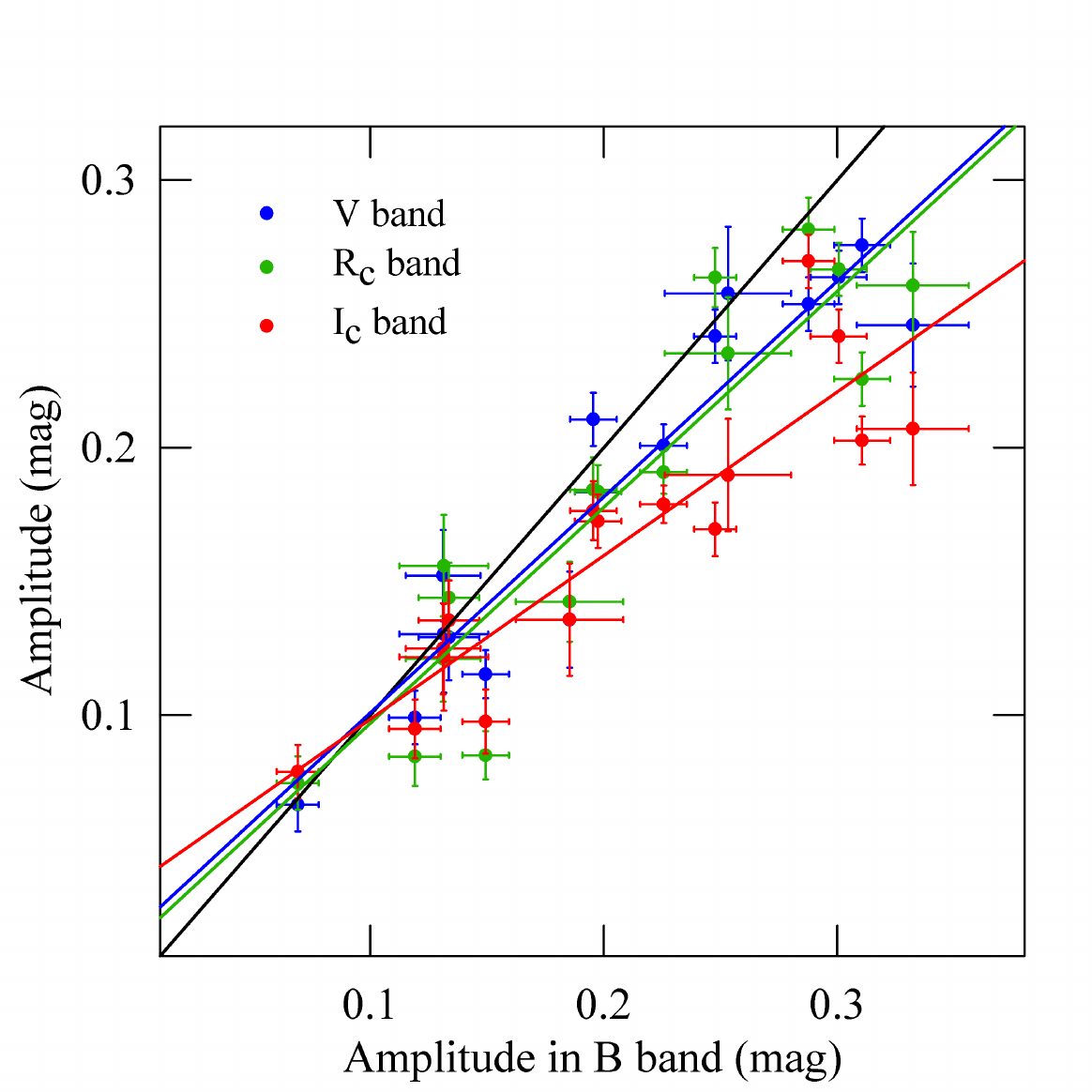}
    \caption{
    Comparison of variability amplitudes in different optical bands.
    The observed data and linear fits for the pairs B and V, B and R$_\text{C}$, B and I$_\text{C}$ are shown in the plot by blue, green, red colors, correspondingly.
    The black line marks the position at which the compared amplitudes are equal.
    }
    \label{fig:amp}
\end{figure}

The variability amplitude was found by the formula \citep{HeidtWag96}
\begin{equation}
A=\sqrt{\left(m_\text{max}-m_\text{min} \right)^2-2\sigma^2},
\label{eq:ampl}    
\end{equation}
where $m_\text{max}$ and $m_\text{min}$ are the maximum and minimum magnitudes of the object during the night, $\sigma$ is the mean error at this night.
The obtained variability amplitudes in the optical bands B, V, R$_\text{c}$, I$_\text{c}$ are listed in Table~\ref{tab:IDVtable}.
From the comparison of variability amplitudes in various bands for nights with the detected IDV (Fig.~\ref{fig:amp}), it follows that amplitudes for different bands are almost equal for small brightness variations (up to $\approx0.15$ of magnitude in the B band). 
For larger brightness variations, the variability amplitude is systematically greater in the filter having a higher effective frequency of the two compared.
This can be explained by the presence of the bluer-when-brighter (BWB) chromatism in IDV.

\begin{figure}
    \includegraphics[scale=0.4]{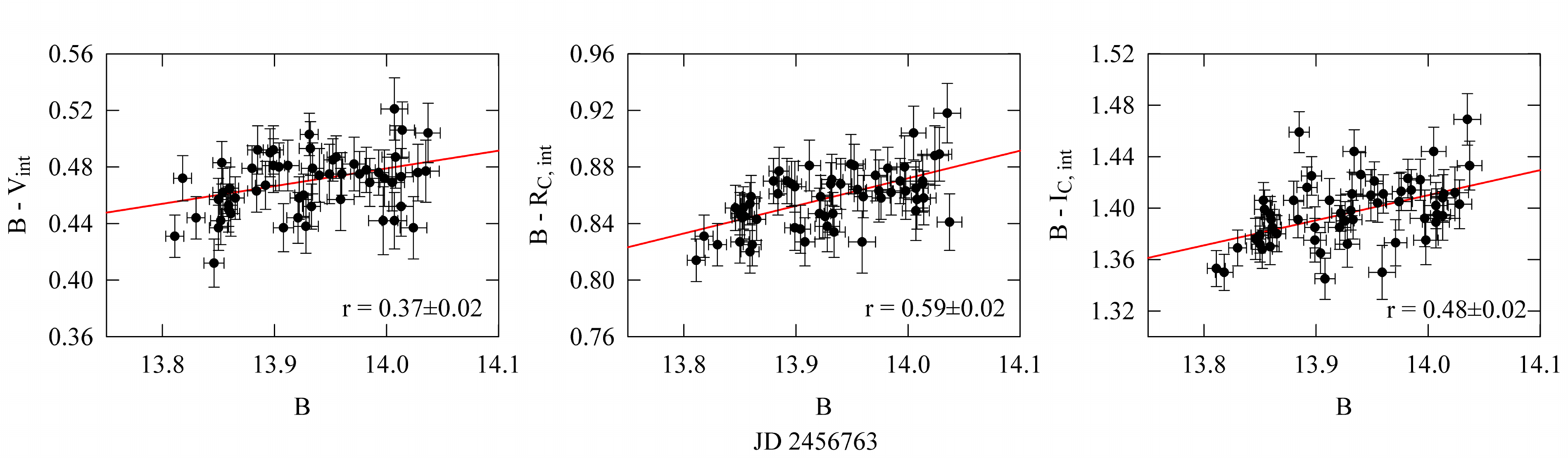}
\caption{The dependence of color indices (B$-$V), (B$-$R$_\text{c}$), (B$-$I$_\text{c}$) on the magnitude in the B band. A solid line shows a linear approximation of the observational data. The corresponding Pearson correlation coefficient $r$ is shown on each plot. There is a trend of increase in r when the difference between effective frequencies of the compared optical bands increases. The full figure is available in Appendix.}
\label{fig:CI}
\end{figure}

The dependence of the color index on the magnitude for nights with the detected IDV is shown in Fig.~\ref{fig:CI}.
The index ``int'' indicates that the magnitude in the corresponding band was linearly interpolated at the moment of observation in the B band.
For the interpolation, we used two adjacent data points in the given optical band. 
The corresponding Pearson correlation coefficient $r$ is shown in each plot.
For most nights, there is a strong BWB trend. And the correlation between the color index and magnitude is slightly higher when comparing optical bands with a larger interval between their effective frequencies.
A similar behavior was also detected by \citet{Dai15}.

\begin{figure}
    \centering
    \includegraphics[scale=0.55]{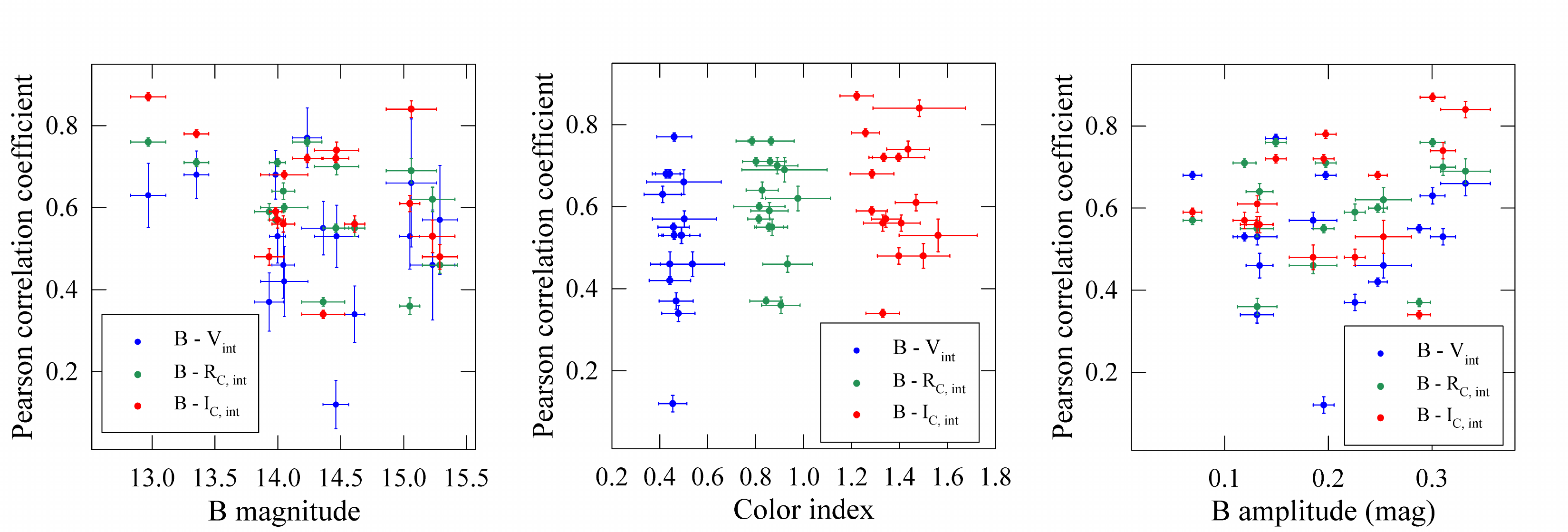}
    \caption{
    The obtained from intra-night data Pearson correlation coefficient ($r$) between the color index and magnitude versus the mean intra-night magnitude (left panel), the color index (middle panel), and the variability amplitude (right panel).
    The error bars in the left and middle plots illustrate the intra-night change in the corresponding value.
    }
    \label{fig:rvsall}
\end{figure}

To reveal conditions for the appearance of intra-day BWB chromatism, we plotted $r$ versus the overnight average magnitude in the B band, the color index B$-$I$_\text{c, int}$, and the variability amplitude for the corresponding night (in Fig.~\ref{fig:rvsall}).
The presence of intra-night BWB chromatism is not clear dependent on the considered parameters.
\citet{Poon09} obtained a similar result.
Based on data over 19 nights of observations on 01-02.2006, \citet{WuZhou07} found that the greater variability amplitude, the bluer color (V$-$R) of the object is observed.
In Sect.~\ref{sec:heljet}, we self-consistently explain the observed different, sometimes contrary, color index behavior during variability.

\begin{longrotatetable}[h]
\begin{deluxetable*}{lllllclll}
\tablecaption{Detailed information for photometric observations of S5~0716+714. Columns contain the following parameters: (1) -- the Julian date; (2), (4) -- F values for the object and comparison star, respectively; (3), (5) -- p-values for the object and comparison star, correspondingly; (6) -- duration of the observation; (7), (8) -- the variability amplitude and its error, respectively;  (9) a label about the IDV detection. Results of the ANOVA-tests, variability amplitudes and their errors are given for filter B, V, R$_\text{c}$ and I$_\text{c}$ in that order.}
\label{tab:IDVtable}
\tablewidth{700pt}
\tabletypesize{\scriptsize}
\tablehead{
\colhead{JD$-$} & \colhead{\shiftright{25pt}{ANOVA for blazar}} & \colhead{}  & \colhead{\shiftright{25pt}{ANOVA for star 18}} & 
\colhead{} & \colhead{\raisebox{-2.5ex}[0cm][0cm]{Duration}} & 
\colhead{\raisebox{-2.5ex}[0cm][0cm]{$A$}} & \colhead{\raisebox{-2.5ex}[0cm][0cm]{$\sigma_\text{A}$}} & 
\colhead{\raisebox{-2.5ex}[0cm][0cm]{IDV}} \\ 
\colhead{2450000} & \colhead{F-value} & \colhead{$p$-value} & \colhead{F-value} & 
\colhead{$p$-value} & \colhead{(hours)} & \colhead{(mag)} &
\colhead{(mag)} & \colhead{} 
} 
\startdata
(1) & (2) & (3) & (4) & (5) & (6) & (7) & (8) & (9) \\ \hline
6751 & 6.54; 3.99; 7.69; 2.65 & 0.0004; 0.007; 0.0001; 0.044 & 0.16; 0.54; 1.26; 1.66 & 0.97; 0.74; 0.31; 0.18 & 2.9 & 0.12; 0.14; 0.11; 0.10 & 0.014; 0.014; 0.014; 0.014 & N \\
6763 & 25.16; 20.12; 20.12; 30.06 & $<0.0001$ & 0.54; 1.91; 1.08; 0.64 & 0.85; 0.08; 0.39; 0.77 & 4.8 & 0.23; 0.20; 0.19; 0.18 & 0.010; 0.008; 0.008; 0.007 & Y \\
6765 & 85.00; 66.02; 68.31; 92.78 & $<0.0001$ & 1.09; 1.72; 0.61; 0.80 & 0.39; 0.11; 0.78; 0.62 & 4.8 & 0.31; 0.28; 0.23; 0.20 & 0.012; 0.010; 0.010; 0.009 & Y \\
6773 & 0.30; 0.54; 2.33; 1.59 & 0.87; 0.70; 0.09; 0.20 & 0.58; 0.55; 0.93; 0.59 & 0.72; 0.74; 0.48; 0.67 & 1.9 & 0.11; 0.16; 0.12; 0.10 & 0.016; 0.017; 0.015; 0.016 & N \\
6778 & 18.07; 18.49; 34.38; 17.75 & $<0.0001$ & 1.93; 1.33; 0.82; 2.33 & 0.07; 0.25; 0.60; 0.03 & 4.0 & 0.25; 0.26; 0.24; 0.19 & 0.027; 0.025; 0.021; 0.021 & Y \\
6780 & 34.95; 23.73; 28.10; 23.85 & $<0.0001$ & 1.24; 0.44; 0.95; 0.55 & 0.28; 0.94; 0.50; 0.87 & 4.4 & 0.33; 0.25; 0.26; 0.21 & 0.024; 0.023; 0.020, 0.021 & Y \\
6781 & 6.15; 7.49; 9.11; 9.66 & $<0.0001$ & 1.56; 0.34; 0.67; 0.56 & 0.17; 0.94; 0.71; 0.80 & 3.2 & 0.13; 0.13; 0.16; 0.12 & 0.019; 0.022; 0.019; 0.020 & Y \\
6785 & 6.35; 9.71; 7.87; 11.23 & $<0.0001$ & 0.95; 0.32; 1.70; 0.71 & 0.51; 0.97; 0.09; 0.71 & 4.4 & 0.13; 0.15; 0.12; 0.12 & 0.016; 0.017; 0.016; 0.017 & Y \\
6794 & 0.52; 0.11; 0.40; 0.38 & 0.61; 0.90; 0.67; 0.69 & 0.16; 0.80; 0.16; 1.12 & 0.85; 0.47; 0.85; 0.35 & 1.0 & 0.04; 0.07; 0.07; 0.12 & 0.017; 0.024; 0.019; 0.024 & N \\
6795 & 4.05; 0.17; 0.74; 3.65 & 0.02; 0.91; 0.54; 0.03 & 1.26; 0.88; 0.10; 0.16 & 0.32; 0.47; 0.96; 0.92 & 1.4 & 0.08; 0.08; 0.07; 0.07 & 0.014; 0.020; 0.017; 0.020 & N \\
6797 & 29.78; 15.78; 23.92; 19.15 & $<0.0001$ & 1.44; 1.81; 1.22; 1.11 & 0.20; 0.10; 0.31; 0.38 & 3.1 & 0.13; 0.13; 0.14; 0.14 & 0.013; 0.016; 0.013; 0.015 & Y \\
6942 & 4.19; 1.54; 7.2; 0.22 & 0.0007; 0.23; 0.0003; 0.92 & 2.41; 0.90; 1.54; 0.32 & 0.07; 0.48; 0.22; 0.86 & 2.0 & 0.04; 0.04; 0.05; 0.04 & 0.008; 0.008; 0.008; 0.010 & N \\
6944 & 2.24; 3.04; 1.86; 2.18 & 0.056; 0.014; 0.12; 0.06 & 1.61; 0.57; 0.64; 0.82 & 0.17; 0.78; 0.70; 0.58 & 2.5 & 0.04; 0.04; 0.04; 0.07 & 0.008; 0.010; 0.009; 0.011 & N \\
6953 & 1.85; 1.82; 1.88; 2.46 & 0.20; 0.20; 0.19; 0.13 & 0.34; 0.18; 1.06; 0.69 & 0.72; 0.84; 0.37; 0.52 & 1.3 & 0.06; 0.06; 0.07; 0.11 & 0.020; 0.017; 0.013; 0.019 & N \\
6959 & 31.58; 14.46; 14.87; 11.60 & $<0.0001$ & 1.60; 1.48; 0.90; 1.36 & 0.11; 0.15; 0.56; 0.21 & 5.4 & 0.15; 0.12; 0.09; 0.10 & 0.010; 0.009; 0.009; 0.012 & Y \\
6966 & 5.23; 15.03; 12.34; 8.46 & $<0.0001$ & 1.10; 1.00; 2.82; 0.85 & 0.38; 0.46; 0.006; 0.85 & 5.8 & 0.19; 0.14; 0.14; 0.14 & 0.023; 0.028; 0.015; 0.021 & Y \\
7072 & 5.93; 4.33; 4.09; 5.38 & $<0.0001$ & 0.96; 1.28; 1.11; 0.95 & 0.54; 0.15; 0.32; 0.56 & 11.1 & 0.07; 0.07; 0.07; 0.08 & 0.009; 0.010; 0.010; 0.010 & Y \\
7074 & 127.43; 96.06; 86.03; 60.35 & $<0.0001$ & 1.30; 1.26; 1.20; 1.20 & 0.15; 0.18; 0.24; 0.23 & 10.0 & 0.25; 0.24; 0.26; 0.17 & 0.009; 0.010; 0.011; 0.010 & Y \\
7075 & 126.40; 102.00; 73.87;& $<0.0001$ & 1.41; 3.81; 0.88; 0.97 & 0.13; $<0.0001$; 0.61;  & 7.9 & 0.20; 0.21; 0.18; 0.18 & 0.009; 0.010; 0.011; 0.010 & Y \\
&  83.40  & & & 0.50 & & & & \\
7076 & 156.19; 171.79; 108.84; & $<0.0001$ & 0.92; 4.15; 1.20; 1.82 & 0.58; $<0.0001$; 0.25 & 11.2 & 0.19; 0.25; 0.28; 0.27 & 0.011; 0.010; 0.012; 0.010 & Y \\
& 130.82 & & & 0.01 & & & & \\
7080 & 14.51; 25.15; 8.08; 11.13 & $<0.0001$ & 0.68; 1.44; 1.25; 0.30 & 0.78; 0.16; 0.26; 0.99 & 6.0 & 0.12; 0.10; 0.08; 0.09 & 0.011; 0.010; 0.011; 0.011 & Y \\
7129 & 130.09; 154.79; 126.06;  & $<0.0001$ & 0.56; 1.02; 0.73; 1.07 & 0.94; 0.45; 0.81; 0.40 & 4.3 & 0.30; 0.26; 0.27; 0.24 & 0.012; 0.010; 0.010; 0.010 & Y \\
 & 108.79 & & & & & & & \\
7130 & 37.87; 36.76; 36.63; 43.40 & $<0.0001$ & 1.07; 1.07; 0.86; 1.11 & 0.39; 0.39; 0.65; 0.35 & 6.6 & 0.20; 0.18; 0.18; 0.17 & 0.010; 0.010; 0.010; 0.010 & Y \\
\enddata
\end{deluxetable*}
\end{longrotatetable}

\subsection{Long-term variability}
\begin{figure}
    \centering
    \includegraphics[scale=0.55]{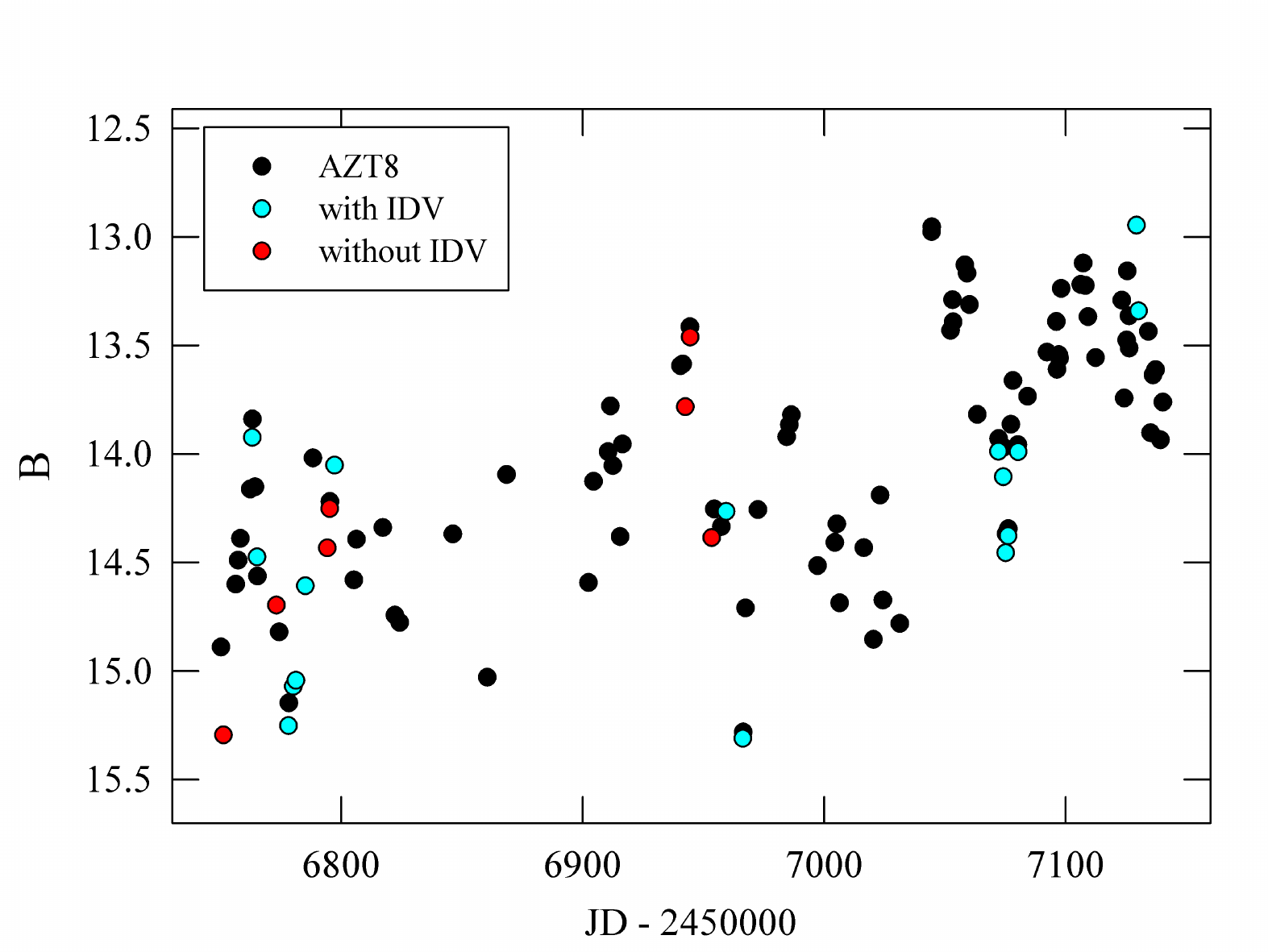}
    \caption{
    Blazar S5~0716+714 light curve for the period 04.2014$-$04.2015 in the B band.
    To plot the light curve, we additionally used data from observations derived with the 70~cm telescope AZT-8 at the Crimean Astrophysical Observatory of RAS (Gorbachev et al. in prep.). Points with and without the detected IDV are represented by cyan and red colors, respectively.
    }
    \label{fig:lc_long-term}
\end{figure}

The long-term light curve of S5~0716+714 during 04.2014$-$04.2015 is shown in Fig.~\ref{fig:lc_long-term}.
To plot the light curve, we completed our data with observational data derived with the 70~cm telescope AZT-8 at the Crimean Astrophysical Observatory of~RAS (Gorbachev et al. in prep.).
We observed the faintest S5~0716+714 on JD~2456751 and JD~2456966, when its mean intra-night magnitude was B=15.31, V=14.78, R$_\text{c}$=14.34, and I$_\text{c}$=13.78 in the optical bands B, V, R$_\text{c}$, and I$_\text{c}$, respectively.
The maximum brightness of the object $\text{B} =12.95$, $\text{V}=12.54$, $\text{R}_\text{c}=12.16$, and $\text{I}_\text{c}=11.74$ was registered on JD~2457129.
These values almost correspond to the faintest and brightest states according to the AZT-8 data (Gorbachev et al. in prep.).
Our IDV observations cover almost the entire range of changes in the object magnitude.
But as seen from Fig.~\ref{fig:lc_long-term}, the occurrence of IDV events does not obviously depend on the object brightness.

\begin{figure}
    \centering
    \includegraphics[scale=0.6]{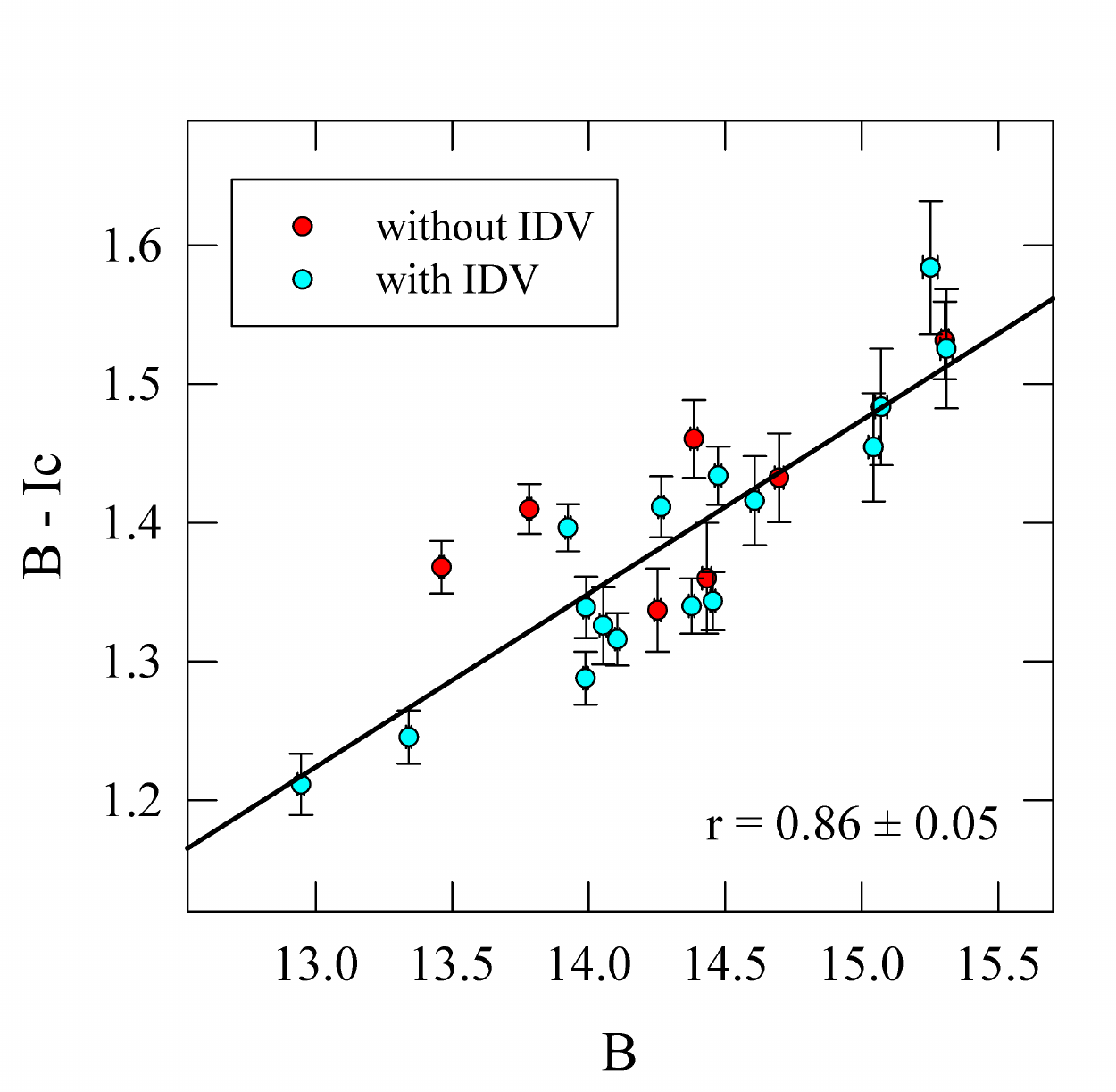}
    \caption{
    Dependence of the color index on the blazar magnitude in the B filter on the long timescale.
    The solid line shows the linear approximation of observational data.
    A strong BWB trend is traced in the data.
    }
    \label{fig:BWB_long-term}
\end{figure}

The long-term variability shows a strong BWB trend: the Pearson correlation coefficient is $r_\text{lt}=0.86$ for all the intra-night averaged data for (B$-$I$_\text{c}$) vs B (Fig.~\ref{fig:BWB_long-term}).
If we consider the nights with and without the detected IDV separately, the long-term BWB trend persists, but the correlation is slightly less ($r_\text{lt}=0.69$) for nights without IDV than for nights with IDV ($r_\text{lt}=0.92$).

\section{Helical jet model in the context of IDV} \label{sec:heljet} 

Hitherto a helical jet in the context of IDV has been considered only during the interpretation of occasionally observed micro-oscillations due to  a helical trajectory of the radiating region located near the jet base \citep{CamenzindKrock92}.
Here we take into account a helical shape of the S5~0716+714 jet on a larger scale: from optical emitting regions to 1~mas from the VLBI-core.
In this framework we assumed that IDV events are produced by sub-components. The sub-component is a part of the main jet component and, during some time, has a Doppler factor significantly higher than that of the main component.
This discrepancy in Doppler factors seems to  be naturally formed by some distinction in motion directions of the main component and its part relative to the line of sight.

\subsection{Geometrical model for IDV} 
\label{sec:geommodel}
\begin{figure}
    \centering
    \includegraphics[scale=0.9]{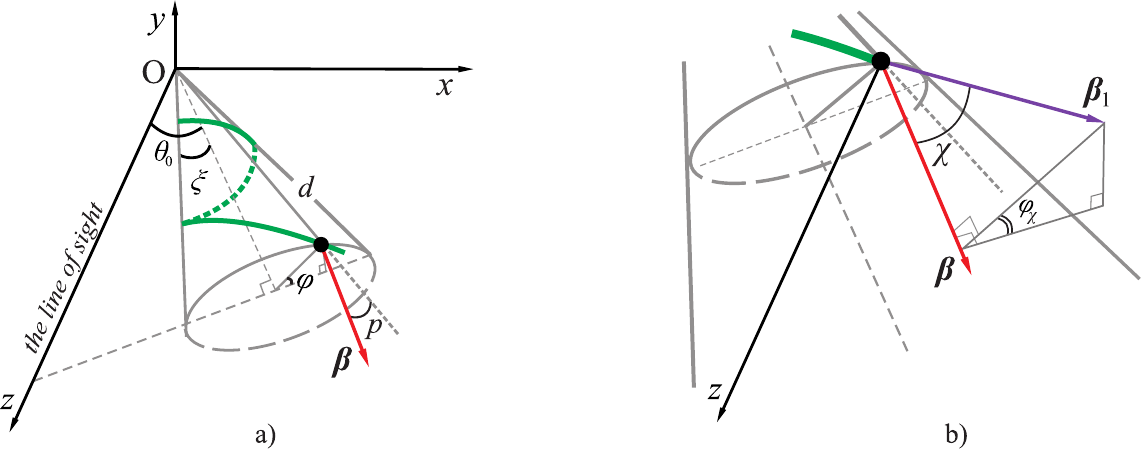}
    \caption{
    a) Schematic of the helical jet. The line connecting individual jet components is marked with green color. Part of the jet located on the side of the notional cone opposite to the observer is denoted by the dotted line. Among all the components, as an example, only one (a black circle) is shown, located at a distance $d$ from the cone apex and moving at the angle $p$ to the cone generating line. The azimuth angle $\varphi$ and the velocity vector \mbox{\boldmath $\beta$} of this component are shown. b) The close-up of the component's vicinity illustrates the motion direction of the sub-component relative to the general trajectory of the main component. To show  a similarity to Fig.~1a in \citep{But18a}, the line of sight Oz is drawn through the jet component.
    }
    \label{fig:heljet}
\end{figure}

In this subsection, we estimate whether the deviation of some part of the jet component from the general trajectory may lead to IDV events.
For this aim, we adopt the helical jet model \citep{But18a, But18b}, which succeeded in providing a mutually agreed description of several observed properties of S5~0716+714.
We briefly recall the main points of this model.
Emitting components of the jet are placed consequently, forming a helical line on the surface of the notional  cone.
We call these components main to emphasize that they are responsible for the long-term properties of the object.
Each of the main components moves at the angle $p$ to the radial direction (Fig.~\ref{fig:heljet}). 
The component position on the helical line relative to the line of sight is described by the azimuth angle $\varphi$.
When the main component moves outwards, its angle $\varphi$ changes, but for a few hours in the observer's reference frame this change is negligible because the distance traveled by the component during this time is much less than that from the cone apex to the component.
The angle between the line of sight and velocity vector, which characterizes the general motion direction of the component, is defined by Formulae (11)–(13) in \citep{But18a}:
\begin{equation}
\begin{split}
 \theta&=\arcsin{\left(\sqrt{f_a^2+f_b^2}  \right)}, \\
 f_a&=\cos p \sin \xi \sin \varphi+\sin p \cos\varphi, \\
 f_b&=\cos p \left(\cos \xi \sin \theta_0+\sin \xi \cos \theta_0 \cos \varphi \right)-\sin p \cos \theta_0 \sin \varphi,
 \label{eq:thnb}   
\end{split}
\end{equation}
where $\xi$ is the half-opening angle of the cone, $\theta_0$ is the angle between the cone axis and the line of sight.
Values of these parameters are obtained by \citet{But18a} and listed in Table~\ref{tab:helparam}. Note, the more arithmetically  compact expression for this angle is obtained by \citet{BP20}.

\begin{table}
 \caption{
The adopted parameters for the helical jet of the blazar S5~0716+714.
 } 
 \label{tab:helparam} 
 \begin{tabular}
 {|c|c|c|}
  \hline
 Parameter & Symbol & Value \\
  \hline
 half-opening angle of the cone & $\xi$ & $1^\circ$   \\
 angle  between the cone axis and the line of sight & $\theta_0$ & $5.3^\circ$ \\
 speed of the components & $\beta$ & $\geqslant0.999$ \\
 deviation of the component trajectory from the radial direction & $p$ & $5.5^\circ$ \\
 \hline
  \end{tabular}
\end{table}

Let us find the angle between the line of sight and velocity vector of the sub-component.
By a sub-component, we mean some part of the component moving with the speed \mbox{\boldmath $\beta_1$} (in units of the speed of light $c$) at the angle $\chi$ to the component's trajectory (Fig.~\ref{fig:heljet}).
Similarly to the angle $\varphi$, we introduce the azimuth angle $\varphi_{\chi}$ characterizing the location of \mbox{\boldmath $\beta_1$} relative to \mbox{\boldmath $\beta$} and the line of sight.
For clarity, the line of sight in Fig.~\ref{fig:heljet}b passes through the jet component.
Figure~\ref{fig:heljet}b is similar to the scheme in Figure~1a in \citep{But18a}.
Then Expression~(5) in \citep{But18a} can be used to find the angle $\theta_\chi$ between \mbox{\boldmath$\beta_1$} and Oz:
\begin{equation}
 \theta_\chi=\arccos{\left(\cos\chi\cos\theta-\sin\chi\sin\theta\cos\varphi_\chi \right)},   
    \label{eq:thhi}
\end{equation}
where $\theta=\theta\left(\xi,\theta_0,\varphi \right)$ is the specified by Formulae~(\ref{eq:thnb}) angle between \mbox{\boldmath$\beta$} and Oz at the considered moment in time.
Formulae~(\ref{eq:thnb}) and (\ref{eq:thhi})  may be used regardless of the physical nature of sub-components.
For example, if the jet main component is a region with a higher number density of emitting particles, then the sub-component is formed by those particles, whose motion direction approximately coincides between themselves and slightly deviates from the general trajectory of the main component.
If the main component is a region containing ejected electrons after their re-acceleration on the shock front, then sub-component forms in that place, where the shock front deviates from its general motion direction.
This deviation may arise due to a small difference in speed of some part of the shock front during shock propagation through the inhomogeneous plasma.

\begin{figure}
    \centering
    \includegraphics[scale=0.3]{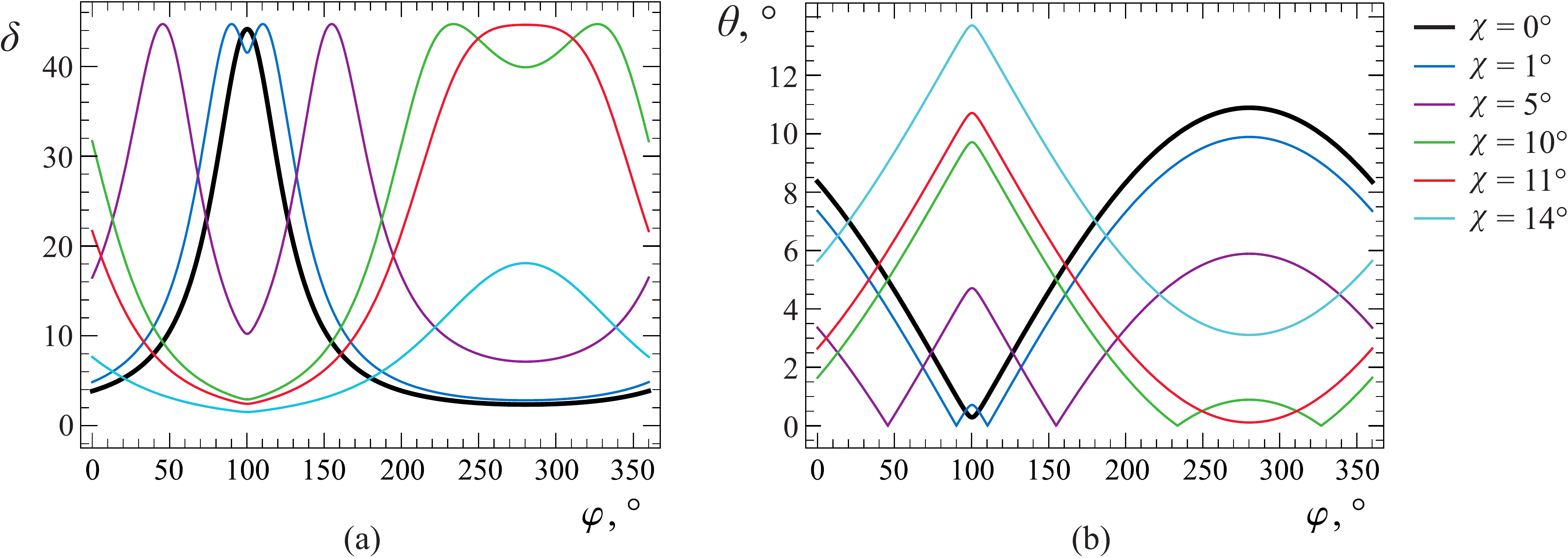}
    \caption{
    a) Doppler factors $\delta$ and $\delta_\chi$ of the jet main component (black bold line) and sub-components (color lines), respectively, versus the azimuth angle of the component.
    The motion of the considered sub-components occurs at angles $\chi$ shown in the plot at $\varphi_\chi=180^\circ$ and $\beta=0.999$; b) The corresponding changes of angles $\theta$ and $\theta_\chi$ between velocity vectors of the main component (black solid line) and sub-components (color lines), respectively.
    }
    \label{fig:dopler_theta}
\end{figure}

In the considered model of the helical jet rotating around its axis, the components consecutively pass the region in which the jet medium becomes transparent for the optical emission. And the angle $\varphi$ of each successive component differs from that of the previous one.
This leads to a change of the Doppler factor
\begin{equation}
\delta=\frac{\sqrt{1-\beta^2}}{1-\beta\cos\theta}.
\label{eq:doppler}    
\end{equation}
The Doppler factor of the main component, located at the distance $d$, versus $\varphi$ is plotted in Fig.~\ref{fig:dopler_theta}a by the black line.
Color lines indicate the Doppler factor $\delta_\chi$ of the sub-component moving at the angle $\chi$ to the trajectory of the component having the azimuth angle $\varphi$.
Figure~\ref{fig:dopler_theta}b shows the corresponding changes of $\theta$ and $\theta_\chi$ versus $\varphi$.
Comparing them, one can see that, e.g., the value of $\theta_\chi$ for $\chi\approx10^\circ$ at $\varphi=200^\circ-360^\circ$ is roughly 20 times smaller than $\theta$, but the corresponding Doppler factor is about 10 times higher. 

\begin{figure}
    \centering
    \includegraphics[scale=0.4]{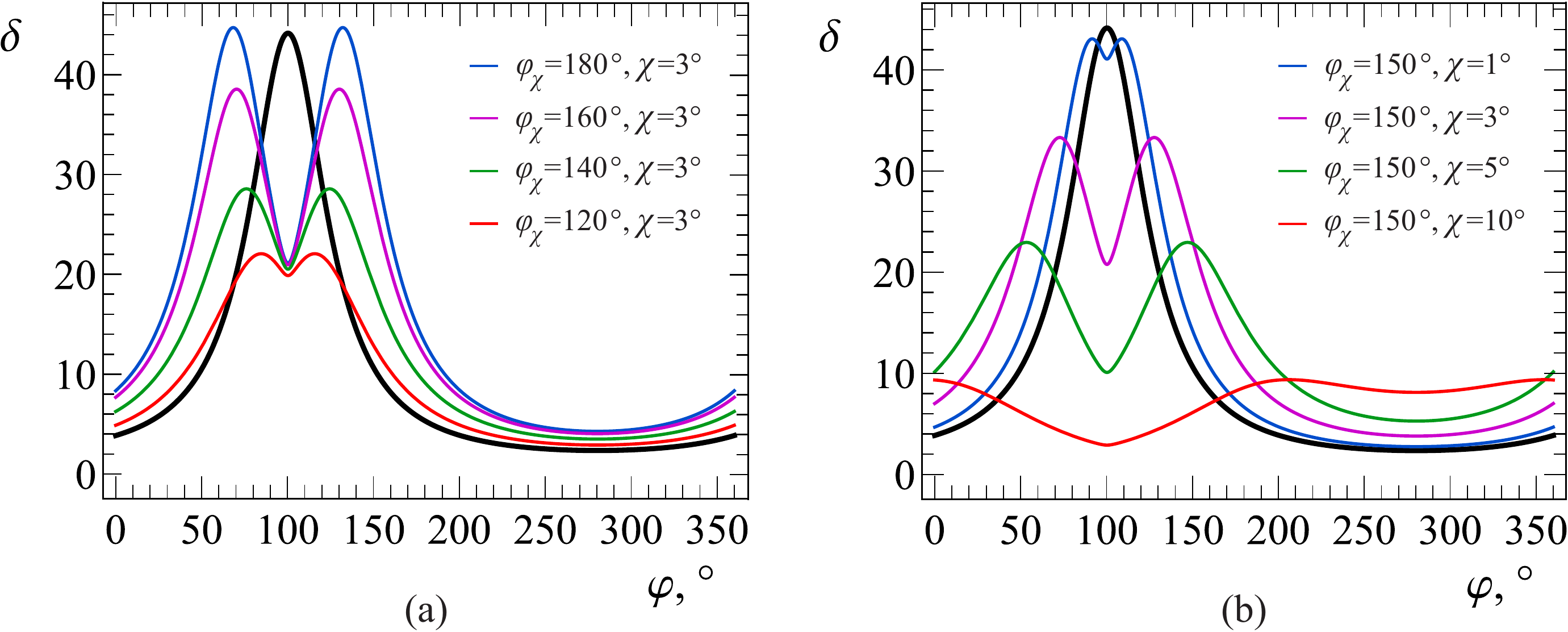}
    \caption{
    Demonstration of the influence of changes in the value of $\varphi_\chi$ on the Doppler factor of the sub-component $\delta_\chi$. (a)~The function $\delta_\chi$ for $\chi=3^\circ$ and $\varphi_\chi$ changing from 180 to 120$^\circ$.
    (b)~The function $\delta_\chi$ for $\varphi_\chi=150^\circ$ and different values of $\chi$.
    For comparison, $\delta$ of the main component is shown by the black line.
    }
    \label{fig:dopler_other}
\end{figure}

\begin{figure}
    \centering
    \includegraphics[scale=0.75]{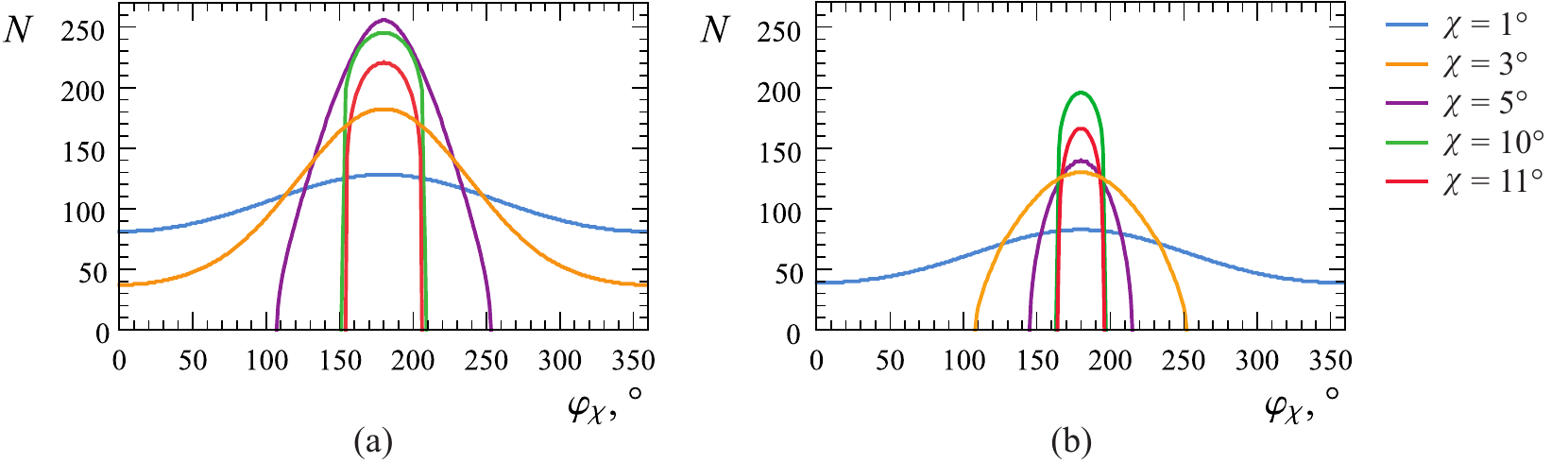}
    \caption{
    The number of values of $\delta_\chi \left( \varphi\right)>10$~(a) and $\delta_\chi\left(\varphi \right)>20$~(b) depending on $\varphi_\chi$ at $\varphi$, varying from 0 to 359$^\circ$ at 1$^\circ$ intervals.
    }
    \label{fig:NfiX}
\end{figure}

As follows from Figure~\ref{fig:dopler_theta}a, at a small difference of $\chi$ from $0^\circ$, the function of $\delta_\chi\left( \varphi\right)$ has two peaks located symmetrically relative to the peak value of the Doppler factor of the main component, $\delta_\text{max}$.
With increasing $\chi$, the difference in the position of these peaks increases: $\varphi$ of the one peak decreases, and that of the other one increases.
For $\chi\approx9^\circ$, both peaks appear in the plot at $\varphi>180^\circ$.  
All the mentioned peaks in $\delta_\chi\left(\varphi \right)$ are almost equal to $\delta_\text{max}$.
With further rise of $\chi$, the value of the single peak of function $\delta_\chi\left(\varphi \right)$ decreases rapidly.
For $\chi\geqslant15^\circ$, the Doppler factor of the sub-component is always $\delta_\chi<10$. 
Thus, for the continuous presence in the radiating region of sub-components with a high Doppler factor  $\left(\approx30-40 \right)$, deviations in their trajectories from the general motion direction of the component up to $11^\circ$ with the constant velocity modulus $\beta=0.999c$ is sufficient.
This conclusion is true for the adopted value of $\varphi_\chi=180^\circ$ under which, as follows from Formulae (\ref{eq:thhi}) and (\ref{eq:doppler}), the value of $\delta_\chi$ is maximum at the fixed other parameters.
As $\varphi_\chi$ decreases, the value of peaks of function $\delta_\chi\left( \varphi\right)$ decreases for the fixed $\chi$  (see Fig.~\ref{fig:dopler_other}a).
Also the value of peaks of the function $\delta_\chi\left(\varphi \right)$ at the fixed $\varphi_\chi$, for example, $120^\circ$, decreases with increasing $\chi$, and the peaks experience a smaller offset along $\varphi$ than at $\varphi_\chi=180^\circ$ (Fig.~\ref{fig:dopler_other}b).
To found the range of values of $\varphi_\chi$, under which $\delta_\chi$ is at least more than 10, we did as follows.
For each value of $\varphi_\chi$ varying from 0 to $359^\circ$ at $1^\circ$ intervals, we estimated the number $N$ of all the values of $\delta_\chi\left( \varphi\right)>10$ when changing $\varphi$ from 0 to $359^\circ$ at $1^\circ$ intervals.
The results are shown in Figure~\ref{fig:NfiX}a. 
The range of values of $\varphi_\chi$, under which the Doppler factor of sub-components reaches 10$-$20, seems to narrow with increasing $\chi$.
Note that the frequent presence of sub-components with $\varphi_\chi=180^\circ$ may not be somewhat peculiar, but perhaps, for example, a consequence of the symmetric distribution of sub-components velocity vectors relative to the \mbox{\boldmath $\beta$}.
Only sub-components with $\varphi_\chi\approx180^\circ$ may be registered as IDV events due to their high Doppler factor.  
Whereas sub-components with other values of $\varphi_\chi$ have a significantly lower Doppler factor and their radiation, although it makes some contribution to the observed radiation of the blazar, but does not lead to IDV events.

\begin{figure}
    \centering
    \includegraphics[scale=0.6]{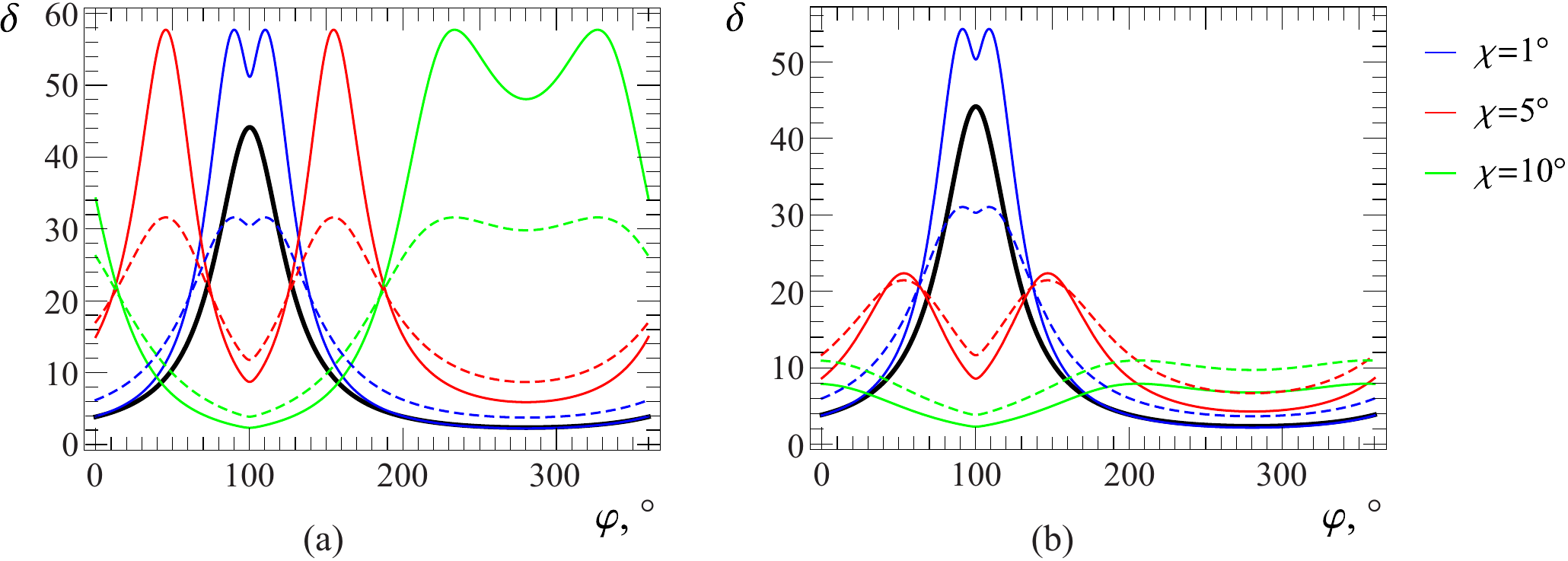}
    \caption{
    The Doppler factor depending on the sub-component speed at $\varphi_\chi=180^\circ$~(a) and at $\varphi_\chi=150^\circ$~(b).
    Color solid lines correspond to $\beta_\text{1,h}=0.9994$, dashed lines --- to $\beta_\text{1, s}=0.998$.
    The Doppler factor of the main component (black line) is shown for comparison. 
    }
    \label{fig:betafiX}
\end{figure}

In the above discussion, we used the lower limit on the speed of main components obtained in \citep{But18a} as a speed of the sub-component $\beta_1$.
Let us consider the variation of $\delta_\chi$ with changing $\beta_1$.  
For instance, we used speeds corresponding to the both $30\%$ increase and decrease in the value of the Lorentz factor for $\beta_1=0.999$.
These values are equal to $\beta_{1\text{,s}}=0.998$ and $\beta_{1\text{,h}}=0.9994$, respectively.
At $\varphi_\chi=180^\circ$ graphs of $\delta_\chi$ maintain their shape, only the value of peaks decreases to $\approx30$ for $\beta_{1\text{,s}}$ and increases to $\approx55$ for $\beta_{1\text{,h}}$ regardless of the value of $\chi$ (for $\chi<11^\circ$).
For small values of $\chi\leq1^\circ$ and, e.g., $\varphi_\chi=150^\circ$, the behavior of $\delta_\chi$ is to be similar to the described above.  
But as $\chi$ increases, the maximum value of $\delta_\chi$ decreases. 
This decrease for $\beta_{1\text{,h}}$ occurs more rapidly than for $\beta_{1\text{,s}}$ (see Fig.~\ref{fig:betafiX}b).   
For $\chi\approx5^\circ$, graphs of $\delta_\chi \left( \varphi\right)$ for various speeds are slightly different. 

Thus, the observed optical radiation of the blazar S5~0716+714 is a superposition of radiation from sub-components and a main component, which have different Doppler factors and different geometrical sizes.
It causes both the continuous strong variability of the object and the absence of prominent maxima on the light curve.
Let us explain it in detail.
Figures~\ref{fig:dopler_theta}, \ref{fig:dopler_other}, and \ref{fig:betafiX} show that the main component has a maximum Doppler factor at $\varphi\approx100^\circ$, while the Doppler factors of sub-components are noticeably smaller.
If the case of the bulk motion of a main component without significant deviations of its parts was fulfilled, this would result in periodic high peaks on the long-term light curve with a low-brightness plateau between them, as observed, e.g., for CTA 102~\citep{Raiteri17, Ammando19}.
For $\varphi\approx0-50^\circ$ and $\varphi\approx150-360^\circ$, the component has $\delta<10^\circ$.
In the latter case, the high brightness of the object may be provided by the presence of sub-components with $\chi\approx8-11^\circ$ and having $\delta_\chi>20$.
Then on the long-term light curve, prominent peaks are absent, and variability is driven by the superposition of individual continuously arising and disappearing sub-components. 
It provides variability with a stochastic behavior noted in \citep{Amir06, Bhatta13}.  
The continuous formation and variation/destruction of sub-components, which at some $\varphi$ are seen at the extremely small ($<1^\circ$) angle to the line of sight, leads to the facts that the object is always active and the possibility of detecting IDV event is high \citep[e.g.][]{WagnerW96, HuChenGuo14, Bhatta16, Agarwal16, HongXiongBai17}.
On the other hand, the presence of sub-components with a lower Doppler factor than that of the main component results in a decrease of the observed flux from the object.
It may occur at $\varphi\approx100^\circ$ when the Doppler factor of the main component is extremely high ($\delta\geq30^\circ$).
As a result, the observed flux from S5~0716+714 does not change as violently as could be expected only from the helical trajectory of components.

\subsection{Do Doppler-factor changes influence on IDV?}
\label{sec:dopler-IDV}

As seen from Figures~\ref{fig:dopler_theta}, \ref{fig:dopler_other}, \ref{fig:betafiX}, Doppler factors of both the main component and sub-components change rapidly in certain intervals of values of $\varphi$.
Let us analyze whether a change of $\delta$ or $\delta_\chi$ caused by variation of $\varphi$ can lead to the geometrical origin of IDV.
For this aim, we consider only changes in $\delta$ because, despite the fact that $\delta_\chi$ depends on several additional parameters, the fastest changes in $\delta$ and $\delta_\chi$ are comparable (Figures~\ref{fig:dopler_theta}, \ref{fig:dopler_other}).

Assuming the flux in the reference frame of emitting plasma is constant, the magnitude change is
\begin{equation}
    \Delta m=-2.5\left(3+\alpha\right)\lg\frac{\delta_1}{\delta_2},
    \label{eq:dm}
\end{equation}
where $\alpha$ is the spectral index of radiation characterized by the power-law spectrum $F_\nu=\delta^{3+\alpha}Q'\nu^{-\alpha}$ ($Q'$ is the proportionality coefficient of the spectral flux in the reference frame of the emitting plasma).
Inserting Formula~(\ref{eq:thnb}) into (\ref{eq:doppler}) and differentiating with respect to $\varphi$, we found that the fastest growth and decrease of $\delta$ occur at $\varphi\approx84.7^\circ$ and $\varphi\approx115.8^\circ$, respectively.
From Equations~(\ref{eq:doppler}) and (\ref{eq:dm}) for $\alpha=1$ it follows that a change in $\delta$ caused by an increase in the mentioned values of $\varphi$ by $1-2^\circ$ leads to $|\Delta m | \approx 0.13-0.27$.
These values are consistent with the observed amplitude of IDV.
Then, if IDV is caused by geometrical effects, the corresponding changes in $\varphi$ should occur in a few hours.
To calculate the time for which $\varphi$ changes by $1-2^\circ$, we use Equation~(10) in \citep{But18a} assuming that the distance of the component from the active nucleus $d$ changes negligibly over $\Delta t$:
\begin{equation}
    \Delta \varphi \approx \frac{\beta c \Delta t \sin p}{d \sin \xi}.
    \label{eq:dfi}
\end{equation}
For $d=4$~pc and $\beta=0.999$ \citep{But18b} during $\Delta t=6$~hours $\Delta \varphi=0.017-0.033^\circ$. Therefore, the Doppler factor changes that lead to the observed variation of magnitude continue over tens of hours.

\begin{figure}
    \centering
    \includegraphics[scale=0.6]{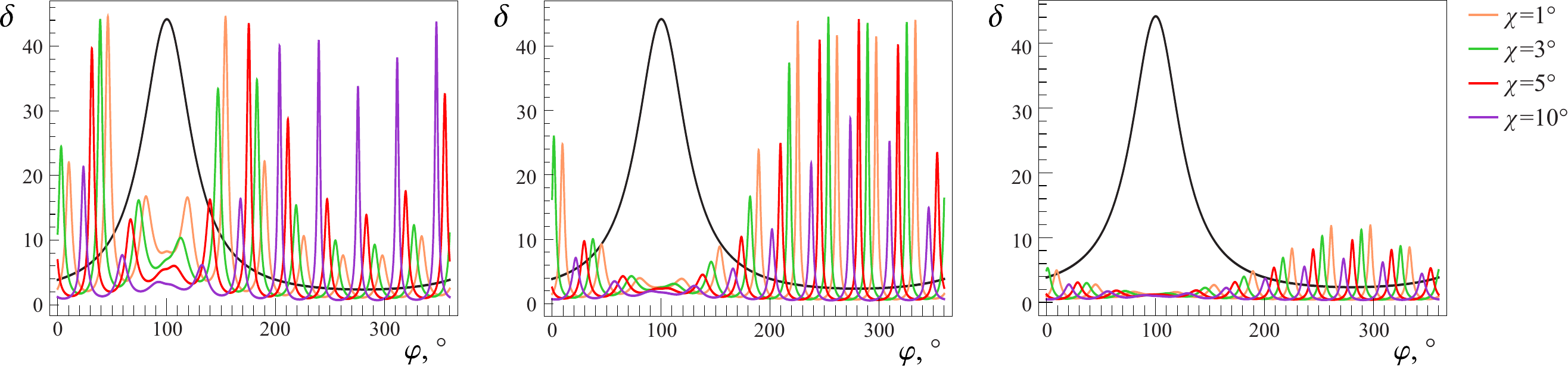}
    \caption{
    Doppler factors of sub-components with $\chi=1$, 3, 5, and 10$^\circ$ rotating around the local jet axis.   
    To plot graphs, we used that the change rate of $\varphi_\chi$ is 10 times faster than the change rate of $\varphi$.
    Each subsequent graph has the initial value of $\varphi_\chi$ that is 9$^\circ$ greater than the previous one.
    The $p_\chi$ values are 5, 10, and 15$^\circ$ for the graphs in the left, middle and right panels, respectively.
    For comparison, the main component's $\delta\left(\varphi \right)$ is shown by a solid black line.
    }
    \label{fig:rotating_subcomp}
\end{figure}

A change in the direction of $\beta_1$ may lead to a change in $\delta_\chi$ on the intra-day timescale.
For example, consider $\delta_\chi$ for the sub-component that rotates around the local jet axis.
This scenario can occur when $n>1$ modes of the Kelvin-Helmholtz instability develop.
These modes produce ``convexities'' on the jet surface that rotate around the jet axis \citep[see, for example,][]{Hardee82}.
By analogy with the non-radial motion described in \citep{But18a}, a change of the angle $\theta_\chi$ between $\beta_1$ of the rotating sub-component and the line of sight is expressed by \autoref{eq:thnb} with the substitution of $\xi$ by $\chi$, $\varphi$ by $\varphi_\chi$, $p$ by $p_\chi$, $\theta_0$ by the angle between the line of sight and the component velocity vector $\theta=\theta\left( \xi, \theta_0, \varphi\right)$ defined by equation~\ref{eq:thnb}.
$p_\chi$ is the angle between motion directions of sub-components with constant and various $\varphi_\chi$ under the fixed $\chi$.
The Doppler factors of sub-components rotating around the jet axis versus $\varphi$ are displayed in Fig.~\ref{fig:rotating_subcomp}.
Thus, we assumed that the change of $\varphi_\chi$ occurs, for example, 10 times faster than the change of $\varphi$ to change $\delta_\chi$ on the intra-day timescale.
Figure~\ref{fig:rotating_subcomp} shows: 1) when $p_\chi$ increases, then peaks of $\delta_\chi$ shift to such values of $\varphi$ at which $\delta$ is small.
2) When $p_\chi$ increases from 5 to $10^\circ$, the values of $\delta_\chi$ peaks decrease for a high $\chi$ and increase for a small $\chi$.
3) When $p_\chi=15^\circ$ for all the considered $\chi$ the maximum $\delta_\chi\leq10$ and $\delta_\chi$ decrease with a further $p_\chi$ growth.
The described above behavior of $\delta_\chi \left(\varphi \right)$ leads to a follow results. The brightness of the main component having the maximum Doppler factor decreases because there are sub-components with a small $\delta_\chi$. While sub-components with a large $\delta_\chi$ increase the brightness of the object at a small $\delta$.
Thus, the object brightness changes within a smaller range than that in the case of bulk motion of the component along the helical trajectory.
Furthermore, the continuous presence, physical and geometrical evolution of sub-components lead to the permanent and strong variability of the blazar S5~0716+714.

\subsection{Color index behavior}
\label{sec:BWB}

As shown in Section~\ref{sec:dopler-IDV}, the sub-component, having the steady flux in the reference frame of the emitting plasma and the constant motion direction relatively to the main component trajectory, produces variability on the timescale of a few days due to the azimuth angle variation of the main component.
Then the following processes can lead to IDV: (i) a physical change in the radiation flux of the sub-component; (ii) a change in the velocity of the sub-component and/or its motion direction relative to the trajectory of the main component.
These factors can operate simultaneously.
If the variability is formed only by the second factor, i.e., the variability is formed by geometrical effects, then there is no dependence of the color index on the object magnitude under the power-law emission spectrum.
\autoref{fig:CI} shows the observed dependencies of color indices B$-$V$_\text{int}$, B$-$R$_\text{int}$, B$-$I$_\text{int}$ on the magnitude in the B band and their linear approximations.
In most cases, there is chromatism.
The BWB chromatism on the long timescale is interpreted in the framework of the shock-in-jet model \citep[e.g. ][]{MarscherG85}.
Then the observed difference between the color indices of the two adjacent IDV events can be explained by the fact that these IDV events are caused by the passage of the shock front through different turbulent cells \citep[see, e.g.][]{Bhatta13, Bhatta16, XuHuWebb19}, which, as \citet{Marscher14} believes, can be present in the jet.
We propose an alternative interpretation of the BWB chromatism in IDV.

In the radio range, most of the radiation detected by a single antenna is originated in the VLBI core \citep{Kovalev05}.
The VLBI core position shifts closer to the true jet base as the observed frequency increases \citep{Pushkarev12}. 
Hence, we can expect a region upstream of the jet where the medium becomes transparent to optical radiation.
By analogy with the radio range, we believe the bulk of the observed optical radiation comes from this region, which we called the optical core.

\begin{figure}
    \centering
    \includegraphics[scale=0.45]{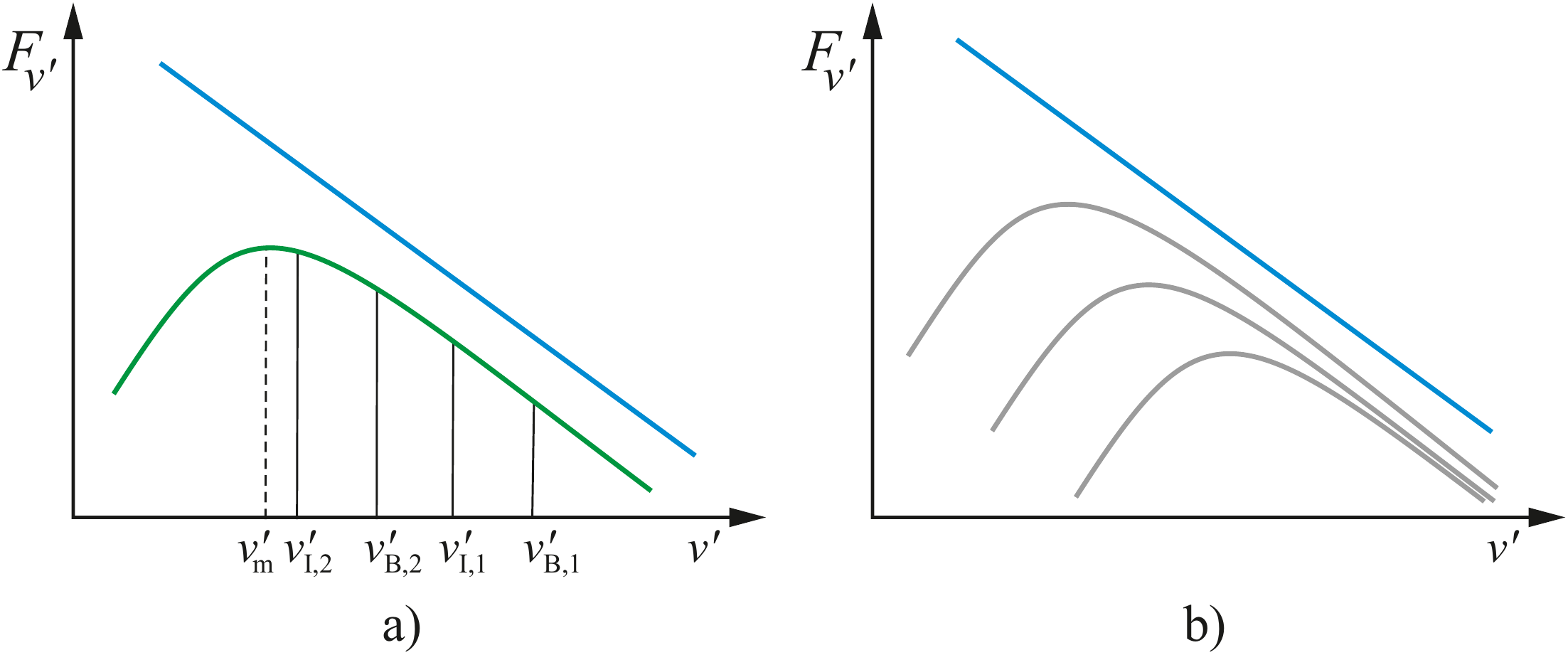}
    \caption{
    a) Scheme for a part of the overall object spectrum (blue line) and the optical core spectrum (green line).
    The frequencies corresponding to the effective frequencies of the B and I optical bands in the minimum (index ``1'') and maximum (index ``2'') brightness of an IDV flare are marked. 
    $\nu^\prime_\text{m}$ is the frequency of the peak flux in the optical core spectrum.
    b) An illustration that the total power-law spectrum of the blazar consists of the sum of the radiation spectra of regions where the jet medium becomes transparent for emission of a certain wavelength.
    }
    \label{fig:SynSAspectrum}
\end{figure}

The spectrum of the optical core has the form schematically represented in Fig.~\ref{fig:SynSAspectrum}a by the green line.
The maximum flux is at the frequency $\nu^\prime_\text{m}=\nu^\prime_1 \tau_\text{m}^{-2/(2 \alpha+5)}$ \citep{Pachol}, where the prime denotes values in the reference frame of the source, $\nu^\prime_1$ is the frequency for which the considered part of the jet has the optical depth $\tau=1$, $\tau_\text{m}$ is the optical depth at the frequency $\nu_\text{m}$, $\alpha$ is the spectral index of the optically thin spectral part.
The overall spectrum of the blazar is formed by the total radiation of regions, in which the jet medium becomes transparent for radiation of different frequencies (Fig.~\ref{fig:SynSAspectrum}b).

The marked in Figure~\ref{fig:SynSAspectrum}a frequencies $\nu^\prime_\text{I, 1}$ and $\nu^\prime_\text{B, 1}$ correspond to the observed effective frequencies of I$_\text{c}$ and B bands before a beginning of IDV flare caused by a sudden change in the Doppler factor of the sub-component $\delta_\chi$.  
At the maximum of the flare, $\delta_\chi$ reaches the highest value that is $\delta_\text{SC}$.
As $\nu=\nu^\prime\delta_\text{SC}$, then the radiation from the sub-component observed in the maximum has the lowest frequencies in the source reference frame.
Frequencies $\nu^\prime_\text{I, 2}$ and $\nu^\prime_\text{B, 2}$ schematically represented in \autoref{fig:SynSAspectrum}a correspond to the observed effective frequencies of the I$_\text{c}$ and B bands for the highest $\delta_\text{SC}$.
In other words, the frequencies $\nu^\prime$, between which the spectral index is determined when the radiation is approximated by the power-law, decrease with an increase of $\delta_\chi$.
This fact and the concave spectral shape will lead to the spectral index decrease with the brightening, i.e., to the BWB chromatism.

\begin{figure}
    \centering
    \includegraphics[scale=0.55]{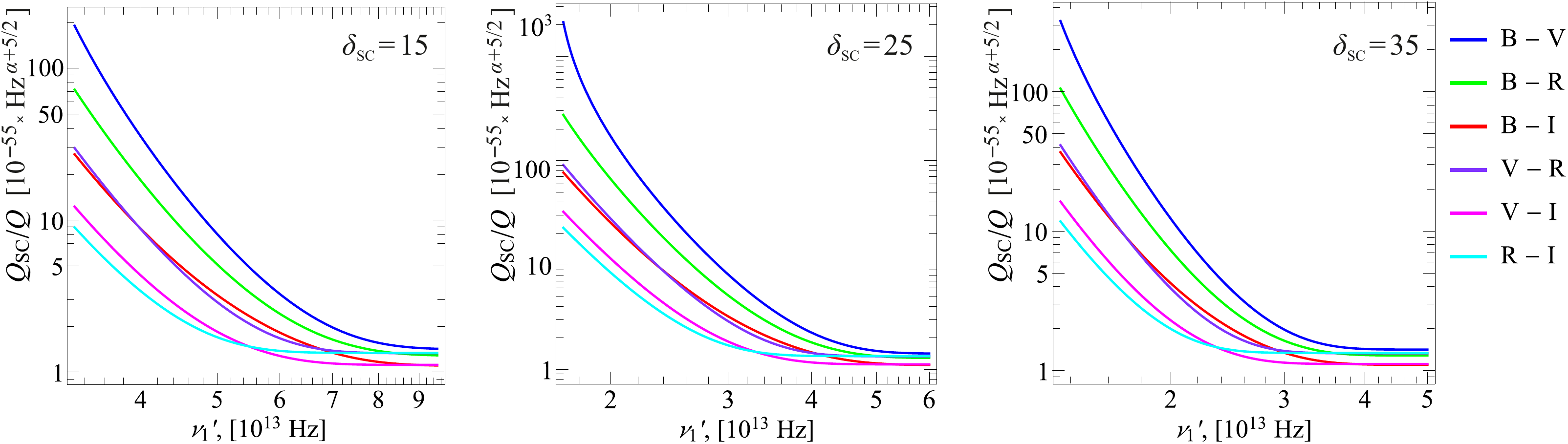}
    \caption{
    Dependence of $Q_\text{SC}/Q$ on $\nu^\prime_1$ for parameters of the IDV flare occurred on JD~2457130.
    Different pairs of the considered optical bands are marked with different colors in the plots.
    The maximum Doppler factor of the sub-component, which responsible for the flare is marked in the upper right corner of each plot.
    }
    \label{fig:57130_BWB}
\end{figure}

Let us examine our assumption about the origin of the BWB chromatism in IDV.
To this aim, consider the IDV flare that occurred on JD~2457130 (Fig.~\ref{fig:lc}).
For this flare, a strong BWB trend was detected, and there are quite a large number of data points (Fig.~\ref{fig:CI}).
Using the calibration of \citep{MeadBallard90}, which associates the magnitude with the flux, the difference in the color index in the bright and low states, for example, between bands B and I$_\text{c}$, is expressed as:
\begin{equation}
   \Delta \left(m_\text{B}-m_\text{I} \right)=2.5 \lg \left(\frac{1+F_\text{B, SC}/F_\text{B, 1}}{1+F_\text{I, SC}/F_\text{I, 1}} \right), 
    \label{eq:dCI}
\end{equation}
where $m$ is the magnitude in the optical band specified in the index, $F$ is the basic spectral flux  (index ``1'') and the flux from the sub-component (index ``SC'') in the corresponding optical bands.
We took the flux corresponding to the minimum brightness of the object for this date (see Table~\ref{tab:SSA-par}) as an underlying basic flux.
When getting expression~\ref{eq:dCI} we assumed the peak flux to consist of the sum of $F_1$ and $F_\text{SC}$ fluxes.
The total emission spectrum, which is the sum of radiation from different regions of the inhomogeneous jet, is well described by the power-law $F_1=Q\nu^{-\alpha}$.
In the observer's reference frame, the spectrum of the sub-component, localized in the region of the optical core, has the form \citep{Pachol}:
\begin{equation}
 F_\text{SC}=Q_\text{SC}\,\nu^{5/2}\left\{ 1-\exp{\left[-\left(\frac{\nu (1+z)}{\delta_\text{SC}\,\nu^\prime_1 }\right)^{-\alpha-5/2}\right]}  \right\} ,  
    \label{eq:Fsc}
\end{equation}
where $Q_\text{SC}$ is the proportionality coefficient, $\nu^\prime_1$ is the frequency in the comoving to the optical core reference frame, $\alpha$ is the spectral index of the optically thin part of the optical core spectrum.
The last parameter cannot be determined from observations, so we assume it equal to the spectral index of the total observed radiation.
Inserting $F_1$ and $F_\text{SC}$ into Formula (\ref{eq:dCI}) we obtained:
\begin{equation}
   \frac{Q_\text{SC}}{Q}=\frac{C-1}{a\left(\nu^\prime_1 \right)-C\cdot b \left( \nu^\prime_1\right)}, 
    \label{eq:ratioQ}
\end{equation}
where $C=10^{0.4 \Delta \left(m_\text{B}-m_\text{I} \right)}$,
$$ a \left(\nu^\prime_1 \right)=\nu_\text{B}^{5/2+\alpha} \left\{ 1-\exp{\left[-\left(\frac{\nu_\text{B}(1+z)}{\delta_\text{SC}\nu^\prime_1} \right)^{-\alpha-5/2} \right]}\right\}, $$
$b\left( \nu^\prime_1\right)$ is expressed similarly to $a\left( \nu^\prime_1\right)$ with the substitution of the effective frequency of the second optical band from the considered ones, in this case it is I$_\text{c}$.
The value of the color index change during variability is determined from the linear approximation of the observed points with the substitution of  magnitudes in the lowest and brightest states (see Table~\ref{tab:SSA-par}).
The dependence of $Q_\text{SC}/Q$ on $\nu^\prime_1$ given by expression~(\ref{eq:ratioQ}) with different values of $\delta_\text{SC}$ for different pairs of optical bands is shown in Fig.~\ref{fig:57130_BWB}.
The area of intersection of curves plotted for different pairs of bands for each $\delta_\text{SC}$ allows us to estimate $\nu^\prime_1$.
The present dispersion of curve intersection points is caused by both the errors in magnitude measurements and the fact that, in neighboring optical bands, the correlation between the color index and the magnitude is weaker than comparing optical bands with a large frequency interval between them (see Table~\ref{tab:SSA-par}).
From Figure~\ref{fig:57130_BWB} it can be seen that for $\delta_\text{SC}=15$ and for Doppler factor of component $\delta=5$, $\nu^\prime_1$ corresponds to effective frequency of the filter I$_\text{c}$. The higher sub-component Doppler factor, the lower $\nu^\prime_1$. 
And $\nu^\prime_1$ for $\delta_\text{SC}=35$ in the observer's reference frame is smaller than $\nu_\text{I}$.
Taking into account the fact that this IDV flare occurs during high state of the blazar (see Fig.~\ref{fig:lc_long-term}), the high Doppler factor of sub-component, producing the observed flare, is expected. 
Thus, there is an agreement between our assumption of the BWB-chromatism and the obtained $\nu^\prime_1=\left(2-4\right)\cdot 10^{13}$~Hz. For $\delta=5$ and $\alpha=1.12$ it follows \citep[see, e.g., ][]{Pachol} that the frequency of the spectrum maximum in the observer's reference frame is $\nu_\text{m}=\left( 0.9-1.7\right)\cdot 10^{14}$~Hz.
The frequency $\nu_\text{m}$ estimated based on the observed data is less than the $\nu_\text{I}=3.7\cdot10^{14}$~Hz.
It corresponds to our assumption that the observed optical radiation comes from the region where the jet medium becomes optically transparent for the considered frequencies.

\begin{table}
 \caption{
Parameters of the BWB-trends of the intra-night flare occurred in JD~2457130, and of the long-term variability. 
 } 
 \label{tab:SSA-par} 
 \begin{tabular}
 {|c|c|c|}
  \hline
 Parameter & Flare in JD 2457130 & Long-term variability \\
  \hline
 Minimum and maximum brightness,  &  &  \\
 respectively (magnitude): & & \\
 B$_1$, B$_2$ & 13.44, 13.24 & 15.31, 12.95 \\
V$_1$, V$_2$ & 13.00, 12.84 & 14.79, 12.55 \\
R$_\text{c,1}$, R$_\text{c,2}$ & 12.61, 12.46 & 14.37, 12.18 \\
 I$_\text{c,1}$, I$_\text{c,2}$ & 12.16, 12.01 & 13.80, 11.73 \\
 \hline
 Spectral index for  & & \\
 minimum brightness state & & \\
 $\alpha$ & 1.12 & 1.46 \\
 \hline
 Linear fits of color indices: & & \\
 $\left( \text{B}-\text{V}_\text{int}\right)$ & $-2.276+0.202\cdot\text{B}$ & $-0.258+0.051\cdot\text{B}$ \\
 $\left( \text{B}-\text{R}_\text{c,int}\right)$ & $-2.505+0.248\cdot\text{B}$ & $-0.197+0.075\cdot\text{B}$ \\
 $\left( \text{B}-\text{I}_\text{c,int}\right)$ & $-2.163+0.256\cdot\text{B}$ & $-0.401+0.125\cdot\text{B}$ \\
 $\left( \text{V}_\text{int}-\text{R}_\text{c,int}\right)$ & $-0.697+0.083\cdot\text{V}$ & $0.057+0.025\cdot\text{V}$ \\
 $\left( \text{V}_\text{int}-\text{I}_\text{c,int}\right)$ & $-0.756+0.123\cdot\text{V}$ & $-0.150+0.077\cdot\text{V}$ \\
 $\left( \text{R}_\text{c,int}-\text{I}_\text{c,int}\right)$ & $-0.400+0.068\cdot\text{R}$ & $-0.201+0.054\cdot\text{R}$ \\
 \hline
 Obtained color index changes: & & \\
 $\Delta \left( \text{B}-\text{V}\right)$ & 0.04 & 0.12 \\
 $\Delta \left( \text{B}-\text{R}_\text{c}\right)$ & 0.05 & 0.18 \\
 $\Delta \left( \text{B}-\text{I}_\text{c}\right)$ & 0.05 & 0.30 \\
 $\Delta \left( \text{V}-\text{R}_\text{c}\right)$ & 0.02 & 0.06 \\
 $\Delta \left( \text{V}-\text{I}_\text{c}\right)$ & 0.02 & 0.17 \\
 $\Delta \left( \text{R}_\text{c}-\text{I}_\text{c}\right)$ & 0.01 & 0.12\\
 \hline
  \end{tabular}
\end{table}

Thus, with an increase of $\delta_\text{SC}$, frequencies in the source reference frame, which correspond to the observed frequencies, decrease.
Different Doppler factors of sub-components cause the frequency shift in the source reference frame to occur by a different value.
The sub-component with a relatively small Doppler factor and a larger occupied volume will result in an IDV flare with achromatic behavior.
Meanwhile, a high $\delta_\text{SC}$ and a smaller volume of the sub-component cause a flare with a similar amplitude with the BWB behavior.
The registered by \citet{Wang19} redder-when-brighter trend can also be explained by the fact that all data for the particular date are used in \citep{Wang19} to plot the dependence of the color index on the magnitude.
But these points can correspond to various features (a monotonous growth or decline of brightness, flares) such as on JD~2458139.
In Section~\ref{sec:obsevid}, we have shown by the example of our data that adjacent IDV events can be characterized by different color behavior in variability.

\begin{figure}
    \centering
    \includegraphics[scale=0.55]{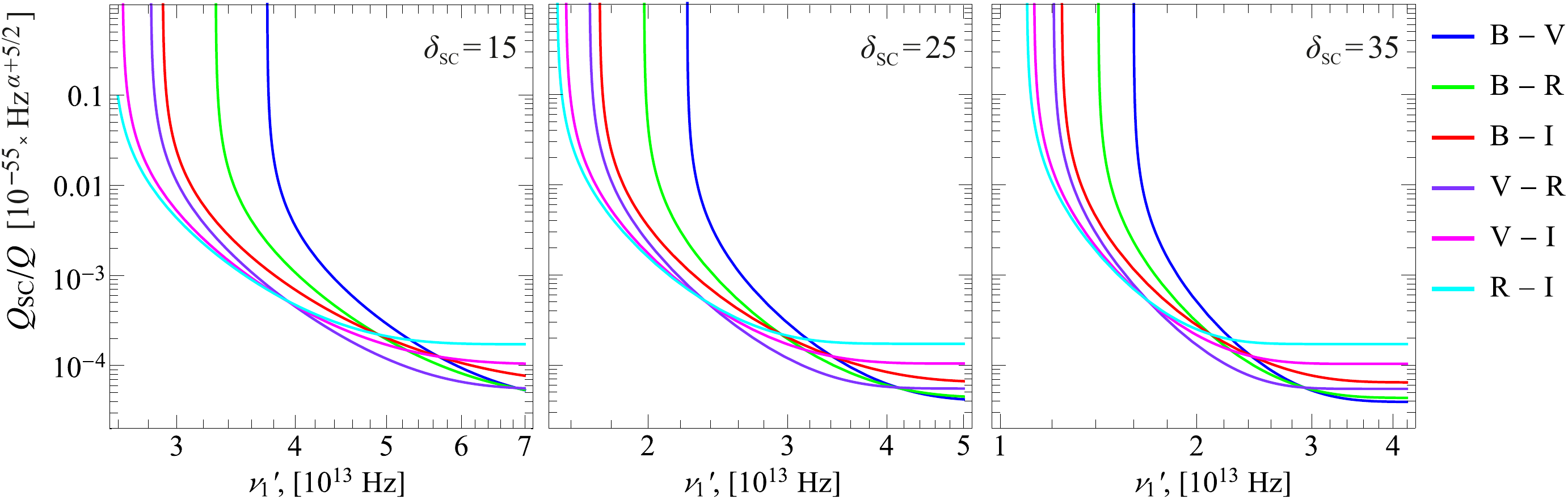}
    \caption{
    Dependence of $Q_\text{SC}/Q$ on $\nu^\prime_1$ for parameters of the long-term blazar S5~0716+714 variability during 04.2014-04.2015.
    Different pairs of the considered optical bands are marked with different colors in the plots.
    The maximum Doppler factor of the sub-component, which responsible for the highest brightness of the object, is marked in the upper right corner of each plot.
    }
    \label{fig:long-term_SSA}
\end{figure}

Similarly to the IDV flare, we consider a long-term BWB trend.
In this case, the variability is formed by not one but many sub-components having different Doppler factors.
We assume that the observed lowest state of the object corresponds to the level of the underlying flux.
An increase in the object's brightness is due to the formation of sub-components with higher Doppler factors.
When the object has the highest brightness, its emission is the sum of radiation of the underlying flux and the flux of the brightest sub-component having the highest brightness due to the highest value of the Doppler factor $\delta_\text{SC}$.
In the same way, as in the previous case, we use linear approximations of the dependencies of the color indices on the object's magnitude to find a change of the color index in variability (Table~\ref{tab:SSA-par}).
The $Q/Q_\text{SC}$ dependencies on $\nu^\prime_1$ obtained for different $\delta_\text{SC}$ are shown in Figure~\ref{fig:long-term_SSA}.

The same as in previous case, the large $\delta_\text{SC}$, the lower $\nu^\prime_1$. 
It seems unlikely that the sub-component with $\delta_\text{SC}=15$ is responsible for the observed variability with an amplitude of about 2.5 magnitudes.
In the cases of $\delta_\text{SC}=25$ and 35, $\nu^\prime_1=\left(1.5-4\right)\cdot 10^{13}$~Hz, which at $\delta=5$ and $\alpha=1.46$ gives the frequency of the spectrum maximum in the observer's reference frame $\nu_\text{m}=\left(0.6-1.6 \right)\cdot 10^{14}$~Hz.
The values of $\nu^\prime_1$ and $\nu_\text{m}$ obtained from the long-term variability are in good agreement with the same values obtained from the particular IDV flare.
This can only be explained by the action of the common variability mechanism on both the long and short timescales, namely the scenario under consideration.

\subsection{Observational evidences}
\label{sec:obsevid}

\begin{figure}
 \centering
\begin{minipage}[h]{0.49\linewidth}
\vspace{4ex}
\center{\includegraphics[width=1\linewidth]{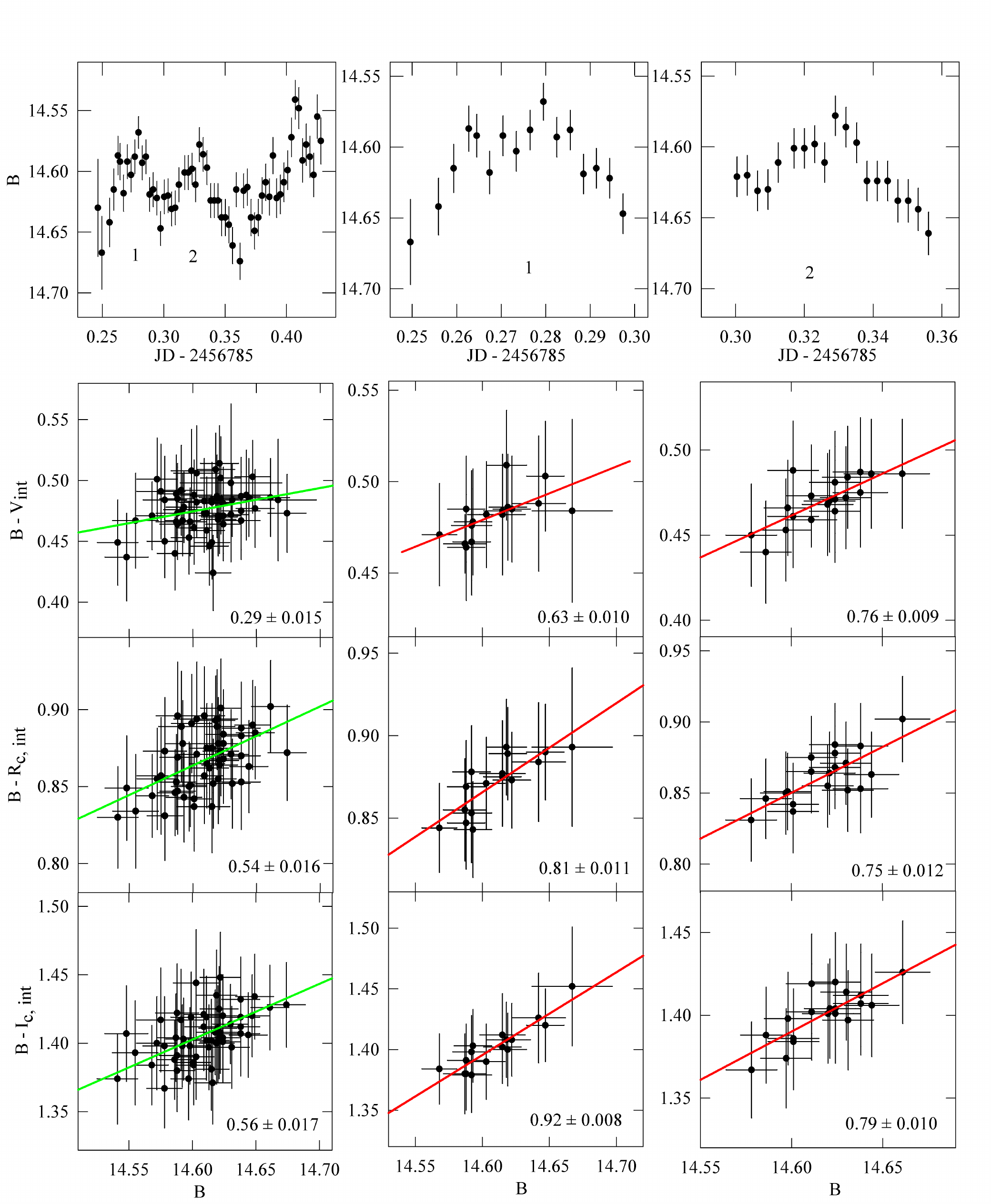} } 
\end{minipage}
\hfill
\begin{minipage}[h]{0.49\linewidth}
\center{\includegraphics[width=1\linewidth]{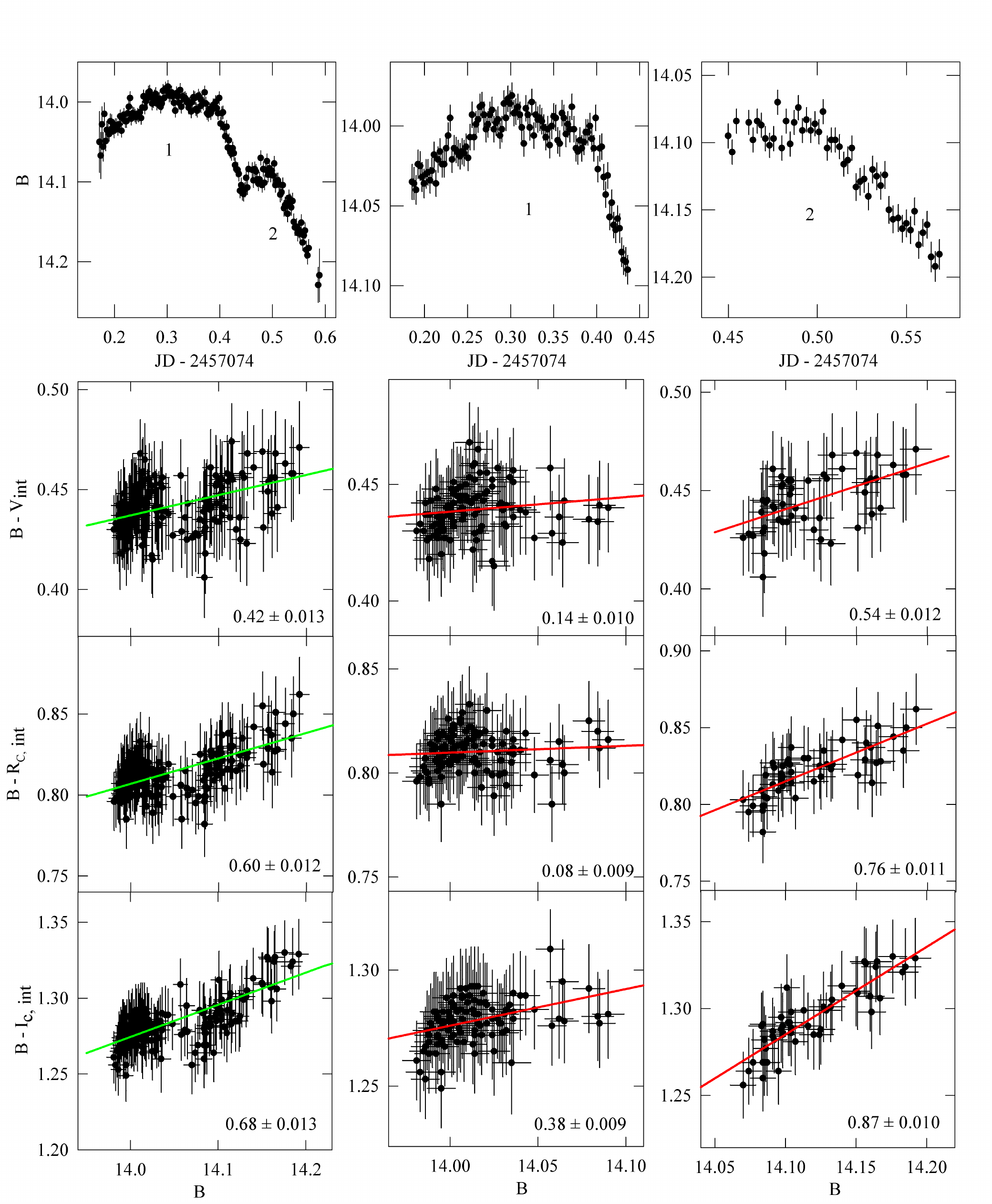}} 
\end{minipage}
\vfill
\begin{minipage}[h]{1\linewidth}
\center{\includegraphics[width=0.49\linewidth]{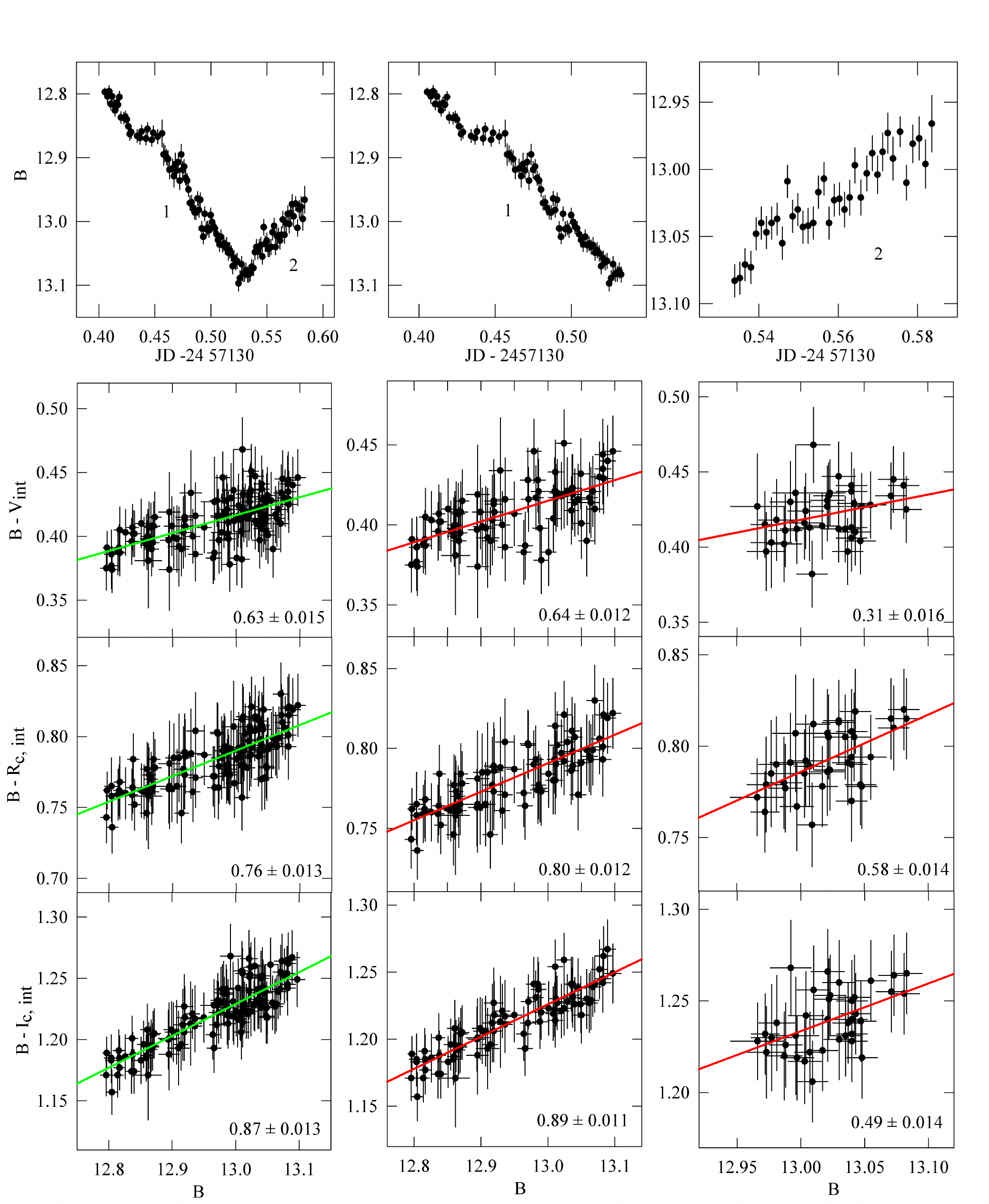} } 
\end{minipage} 

\caption{The intra-night light curve in the B band (top) and the color behavior in variability (bottom) for JD~2456785 (top left), JD~2457074 (top right), and JD~2457129 (bottom). The linear approximation and the Pearson correlation coefficient are given for each dependence of the color index on the magnitude. All intra-day data are shown in the left panel, data for individual parts, in which the total light curve was divided for each night, are shown in the middle and right panels.}
\label{fig:obsevid}
\end{figure}

Our assumption about the origin of IDV can be confirmed by the fact that different features on the intra-day light curve have different dependencies of the color index on the magnitude.
This possibility was mentioned by \citet{Amir06, Wang19}, but was not investigated in detail.
For example, two adjacent flares of approximately the same amplitude and duration were observed on JD~2456785 (Fig.~\ref{fig:lc}).
The light curve and dependence of the color index on the magnitude are shown in Figure~\ref{fig:obsevid}.
The color index obtained from all data for the night shows a weak correlation with the object brightness, the Pearson correlation coefficient $r\approx0.5$.
Meanwhile, there is a stronger correlation between the color index and the magnitude under separate consideration of these two flares.
This observational result can be interpreted by the fact that if the flares are formed by a changing Doppler factor of the corresponding sub-components, the values of the maximum Doppler factors of sub-components differ.
The difference in the value of $\delta_\text{SC}$ with the concave optical spectrum of the region from which the observed radiation comes may lead to various changes in the color index of the object in variability.

For flares that occurred on JD~2457074, the correlation coefficient between the color index and the magnitude is high Fig.~\ref{fig:obsevid}.
But separately for the first flare $r\approx0.1$ for B$-$V$_\text{c}$ and B$-$R$_\text{c}$, and slightly higher ($r\approx0.4$) for B$-$I$_\text{c}$.
Whereas for the second flare $r$ reaches 0.87 for B$-$I$_\text{c}$.
The absence of BWB for the first flare can be explained by the fact that either the flare is a superposition of several flares with different color index, or the flare is formed by a relatively small increase in the Doppler factor of the sub-component, which does not change the color index under the concave optical spectrum.

In JD~2457129, a decrease in brightness at a rate of 0.096~mag~h$^{-1}$ replaced an increase in brightness at about the same rate.
For all the data, the correlation between the color index and brightness is strong ($r\approx0.8$).
Under separate consideration, a strong correlation is maintained for the brightness decline and becomes less ($r\approx0.5$) during an increase in brightness.

\subsection{Magnetic field estimation}

\begin{figure}
    \centering
    \includegraphics[scale=0.85]{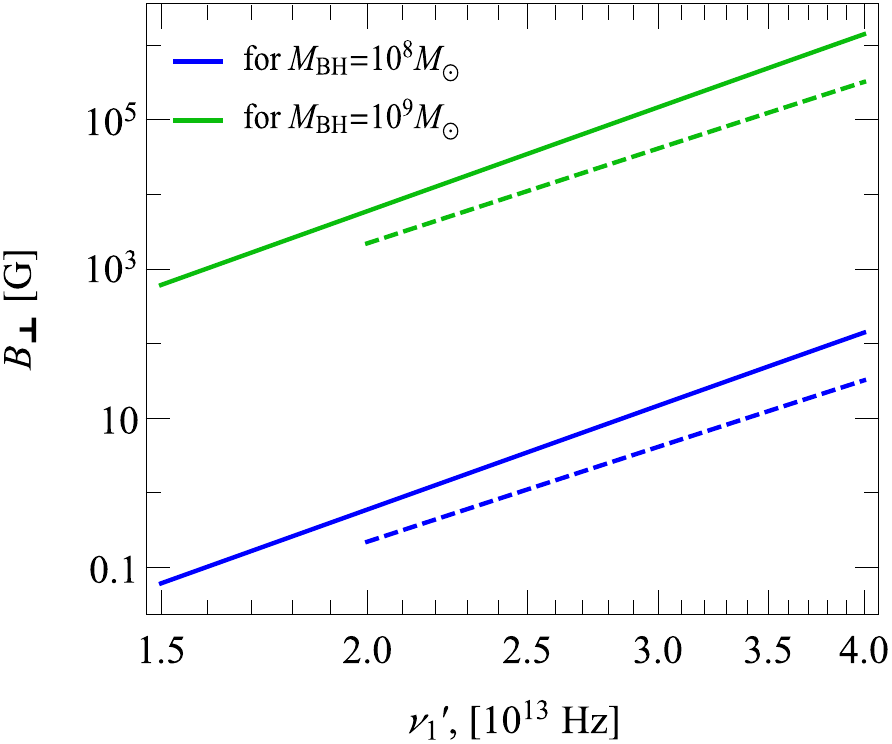}
    \caption{
    Dependence of the magnetic field in the optical emitting region on $\nu^\prime_1$ for parameters of the JD~2557130 IDV flare (dashed lines) and the long-term blazar S5~0716+714 variability during 04.2014-04.2015 (solid lines). 
  Sizes of the emitting region are assumed to be equal to the gravitation radii of black holes with masses of $10^8$ and $10^9$ solar masses. 
    }
    \label{fig:SSAmf}
\end{figure}

If the turn-over frequency of the synchrotron spectrum and the source size are known, it is possible to estimate the transverse to the line of sight component of a magnetic field $B_\perp$ \citep{Slish63}.
The well-known formula for $B_\perp$ (see, e.g., \citep{Marscher83} and detailed derivation in \citep{PushkarevBKH19}) uses the observed angular size of the source, which is unknown in our case. 
We assume that the size of the optical radiating region in the reference frame, comoving with the relativistically moving plasma, is comparable to the gravitational radius of a central black hole $R^\prime=R_g$. 
Using relativistic transformations (see Appendix C in \citep{BBR84})  and cosmological reduction of the observed spectral flux, we obtained 
\begin{equation}
F_\nu=\left( 1+z \right)^{-3} \delta \cdot I^\prime_{\nu^\prime}\left( \nu^\prime\right)d\Omega^\prime,
\label{eq:Fobs}
\end{equation}
where $d\Omega^\prime=\pi R^{\prime\,2}/(4 D^2_L)$ is the solid angle of the source in the comoving with relativistically moving plasma reference frame, $D_L$ is the luminosity distance of the object, the special intensity is \citep{Pachol}
\begin{equation}
    I^\prime_{\nu^\prime}\left(\nu^\prime \right)=\frac{c_5}{c_6}B_\perp^{\prime\,-1/2} \left(2 c_1 \right)^{-5/2} \nu^{\prime\,5/2} \left[1-\exp\left(- \tau_\nu \right) \right],
    \label{eq:InuS}
\end{equation}
where $c_1$, $c_5$, $c_6$ are the constant and functions tabulated in \citep{Pachol}, $\tau_\nu=\left(\nu^\prime/\nu^\prime_1 \right)^{-\alpha-5/2}$ is the Lorentz invariant optical depth at the frequency $\nu$, $\alpha$ is a spectral index of optically thin part of the spectrum. Combining the expressions (\ref{eq:Fobs}), (\ref{eq:InuS}) and taking into account $\nu^\prime=\nu\left(1+z\right)/\delta$, we obtained the perpendicular to the line of sight component of the magnetic field in the source reference frame 
\begin{equation}
    B^\prime_\perp=\left\{ \frac{\pi}{4} \frac{c_5}{c_6}\left[ 1-\exp\left(-\tau_\nu \right)\right]\right\}^2 \left( 2 c_1 \right)^{-5} \frac{R^{\prime\,4}}{D_L^4} F_\nu^{-2} \nu^5
\left( 1+z\right)^{-1} \delta^{-3}.   
\label{eq:Bperp}
\end{equation}
Using in formula (\ref{eq:Bperp}) the power-law approximation of the observed fluxes in different filters for the effective frequency, e.g., of filter I$_\text{c}$, for IDV flare (JD~2457130) and long-term variability, we plotted the magnetic field versus $\nu^\prime_1$ (Fig.~\ref{fig:SSAmf}). 
It can be seen that for the size of the radiating region, which is equal to the gravitational radius of a black hole with a mass of $5\cdot10^8$ solar masses (as expected for blazars in \citep{SbarratoGM12, Titarchuk17} and references therein), the magnetic field strength is within the values estimated by other authors from independent models \citep[e.g.,][and references therein]{FieldR93,Mondal19}.

\section{Discussion}
\label{sec:discussion}
The absence of any features in the non-thermal spectrum of S5~0716+714 \citep{Stickel93} and unsuccessful attempts to resolve a host galaxy of the blazar \citep{Scarpa00, Sbarufatti05} indicate that most of the observed radiation from the object is formed in its relativistic parsec-scale jet.
This radiation, undergoing the Doppler boosting, shadows the radiation originating from the rest parts of the active galaxy.
Therefore, models connecting IDV to the accretion disk \citep[e. g., ][]{MangalamWiita93, Sun14, Gupta09, Hong18} are not applicable for this source.

A helical jet has been used for about 30 years to interpret different properties of active galactic nuclei.
For example, \citet{CamenzindKrock92} explained the quasi-periodic flares of brightness observed on the long timescale for objects 3C~273 and BL~Lac by that the radiating plasma moves along helical trajectories near the magnetized accretion disk, thus having a variable Doppler factor.
The jet can acquire the helical shape as a result of the jet nozzle precession.
Then the precession period of several years can only be achieved in the binary black hole system, in which the accretion disk of the primary does not coincide with the plane of the companion's orbit \citep[e.g., see one of the first papers][]{Katz97}.
This scenario was considered for many objects, for example, for S5~0716+714 \citep{Nesci05} and OJ~287 \citep{Abraham00}, and allowed estimating the masses of black holes \citep{LiuWu02} from quasi-periods of long-term variability.
As \citet{But18b} noted, the non-radial motion of the jet main components leads to an increase in the period of long-term variability in the radio range \citep[5$-$6 years for S5~0716+714, ][]{Raiteri03, Gupta08, Tang12} compared to the period of variability in the optical range \citep[3.0$-$3.5 years, ][]{Raiteri03, Liu12, Bychkova15, LiLuo18}. Therefore, the estimations of the orbital periods and masses of black holes, especially those derived from the analysis of data in the radio range \citep[for example, ][]{Bychkova15},  are distorted. 
Moreover, recently for OJ~287, it was shown that observed periods of optical flux variations and changes of inner jet position angles were explained by the helical jet with non-radial component motion, which can be formed by Kelvin-Helmholtz instability. Hence, the assumption of binary black hole is needless \citep{BP20}.

An alternative scenario for the origination of the helical jet is the development of (magneto-) hydrodynamic instabilities \citep{SteffenZensus95}, among which the Kelvin-Helmholtz instability is widely discussed \citep[e.g., ][]{Hardee82, Hardee03}.
The advantage of (magneto-) hydrodynamic models over precession models is that the former have non-radial plasma motion in the jet \citep{Hardee82, Hardee03, Nokhrina19, MertensLobanov16}.
The motion along the curved paths was detected for features  of many parsec-scale jets \citep{Lister16, Lister19}, including S5~0716+714 \citep{Rast09, Rast11, Rani15}.
The motion of components of the S5~0716+714 blazar jet at an angle of $5.5^\circ$ to the radial direction allowed us to provide a consistent explanation of the different observed facts \citep{But18a, But18b}.
It is noteworthy that the value of the azimuthal component of the speed of the S5~0716+714 jet component is consistent with the value obtained when modeling the shape of the M~87 parsec-scale jet \citep{Nokhrina19, MertensLobanov16}  and related to its inner part (up to a deprojected distance of 20~pc).
The agreement of azimuthal speeds of the jet parts provides indirect evidence that the geometric and kinematic model of the S5~0716+714 jet used by us is appropriate.

For S5~0716+714, there are time delays in long-term variability at a lower frequency relative to variability at a higher frequency \citep{Raiteri03, Fuhrmann08, Rani13, Rani14, LiLuo18}.
This indicates that the regions in which the the radiation at different frequencies generates are spatially separated.
The higher the frequency of radiation coming from a certain region, the closer to the true jet base this region is located.
In the radio range, this ``separation'' of radiating regions naturally occurs due to the action of synchrotron self-absorption \citep{Konigl81, Lobanov98, Pushkarev12}.
In the spectral energy distribution of S5~0716+714 the low-frequency synchrotron bump extends up to X-ray range \citep[e.g., ][]{Liao14}, thus the optical radiation is generated by the synchrotron mechanism.
Therefore, we can expect a region in the jet where the medium becomes transparent to optical radiation.
In the model of the relativistic Blandford-K{\"o}nigl supersonic jet \citep{BlandfordKonigl79}, for the optical radiation a fast increase of transparency with distance from the true jet base is expected: $r\propto \nu^{-1/k_{r}}$ \citep{Konigl81}, where $k_{r}\approx1$ \citep{Lobanov98}.
Therefore, if the 15~GHz VLBI core is located at a distance of $r_\text{15 GHz}=6.68$~pc from the true jet base \citep{Pushkarev12}, then the optical core is located at a distance of $r_\text{Opt}=r_\text{15 GHz}1.5\cdot 10^{10}/\left(4.5\cdot10^{14}\right)\approx2.2\cdot10^{-4}$~pc from the jet beginning (i.e. from the location where flow from the active nucleus have been collimated into supersonic jet, described by \citet{BlandfordKonigl79}).
On the other hand, it was estimated the distance of the optical core from the apex of the notional cone, on the surface of which the S5~0716+714 helical jet is located, and it is 4.6~pc \citep{But18b}.
The considered conical shape is shown in stacked multi-epoch maps of jets, including S5~0716+714 \citep{Pushkarev17}.
Due to the sufficient linear resolution for the nearest jets, a parabolic shape was detected at the stacked maps nearby to the true jet base \citep{Pushkarev17, Kovalev19}.
The transition from the parabolic to conical shape occurs at a distance of $10^5-10^6$ gravitational radii from the core.
This transition may be a common property of jets \citep{BeskinNokhrina06, Beskin17}, and, if it also occurs in the S5~0716+714 jet, the active nucleus may be significantly closer to the radiating region in the optical range than the apex of the cone considered by \citet{But18b}.
The question about the distance of the optical core from the black hole remains open. But this does not qualitatively affect the conclusions of our IDV study.
Thus, if the optical core is located at a smaller distance from the true jet base than the distance at which the transition of the stacked shape of the jet from parabolic to conical occurs, then we can consider the axis of the helical jet located on the surface of the paraboloid of revolution.
The optical core, being at a constant distance from the paraboloid apex, periodically changes the azimuth angle due to the jet's outward motion.
The motion of the optical core on the surface of the paraboloid can be described by the motion on the surface of the cone formed by tangents to the paraboloid at each point of the trajectory of the optical core.
Then, the presented in Section~\ref{sec:heljet} formulae can be applied with taking into account that the closer the optical core to the paraboloid apex, the greater the cone opening angle.

Studies of the color index change of S5~0716+714 in variability give different results.
Our IDV observations, as well as those of \citep{Amir06, HuChenGuo14, Ghisellini97, WuZhou07, Bhatta16, Dai13, Wu05, Stalin09, Feng20}, show a strong BWB trend.
We and other researchers \citep{HuChenGuo14, WuZhou07, Dai13, Kaur18, Stalin09} note the BWB trend on a long timescale, while authors of \citep{Ghisellini97, Raiteri03} report the achromatic behavior of long-term variability.
The absence of chromatism in both IDV and long-term variability was claimed in \citep{Stalin06, Agarwal16}. 
Both the achromatic behavior and the BWB trend in IDV were found by \citet{HongXiongBai17, Raiteri03}.
At some nights, it was found a weak redder-when-brighter (RWB) trend \citep{HongXiongBai17, ZhangWM18, Ghisellini97}, but we believe that it arose from the fact that different IDV features  on the light curve were considered together.
In other words, one variable component is replaced by another.
For example, a sub-component, having a large Doppler factor and a small occupied volume, makes the color index bluer for an intermediate brightness level. 
Further replacement of this sub-component with another, having a smaller Doppler factor and a significantly larger occupied volume, leads to the brightness increase and reddening of the object's color. 
This scenario works for BL Lac objects, which have featureless optical spectra corresponding to the negligible contribution of accretion disk emission to the total flux. 
Flat Spectrum Radio Quasars, for which the significant thermal emission from the accretion disk is detected, show substantial redder-when-brighter and bluer-when-brighter trend. 
An explanation of these behaviors is that there are two variability components: ``blue'' and ``red'' \citep{Agarwal16,GuLee06,Isler17}. 
The red component is the synchrotron radiation of the jet, while the blue component is the thermal radiation of the accretion disk.
When FSRQ is brightening due to increase flux from the jet, the RWB trend is observed. 
But accounting for radiation from the vicinity of the central engine or the host galaxy, as we mentioned above, is not applicable for S5~0716+714.

The color index behavior during variability allows us to conclude about the mechanisms leading to the brightness changes of the blazar.
Namely, geometrical effects under the steady power-law spectrum of radiation do not change the object color index in variability \citep{G-KW92, CamenzindKrock92}. Meanwhile, physical processes, for example, the shock passage in the jet plasma \citep{Kirk98} or the injection of high-energy particles into the radiating region \citep{Ghisellini97}, change the color index.
The passage of the shock downstream of the jet produces variability on the timescales from weeks to a few months.
However, IDV events can be caused by shocks propagating in the turbulent jet, and optical radiation must be generated by inverse Compton scattering, e.g., \citep{KoniglCh85, Hughes89}.
\citet{Marscher14} proposed a model (TEMZ) in which a turbulent jet passes through the standing conical shock.
Optical radiation is assumed to be synchrotron.
Although one of the advantages of the TEMZ model is that it explains the observed rotation of the electric vector in the wave, but on a time scale of several hours, the changes in the electric vector direction in the wave observed for S5~0716+714 correspond to the motion of a particle in the precessing helical magnetic field \citep{ShabAf19}.
The TEMZ model was applied to interpret the IDV flares of the blazar S5~0716+714 \citep{Bhatta13, XuHuWebb19}.
The authors estimated the size of the turbulence cells to be 5$-$166~au and showed the expected time lag of variability between the V, R, and I bands to be consistent with the observed one. Unfortunately they made no estimates and conclusions about the color index behavior during the flares, and they did not take into account a different angle between the velocity vector of the considered cells and the line of sight.

The possibility of the IDV event origin due to geometrical effects was often considered by different researchers.
For example, \citet{Bachev12} supposed that the trajectory of some relativistically moving blob may slightly deviate from the straight line resulting in a slightly increased Doppler factor for this particular blob.
On the other hand, the formation and rotation of the warp inner part of the accretion disk lead to small perturbations of the relativistic electron-positron flow and, consequently, to the intra-day variability of the beamed flux caused by a change in the angle with the line of sight \citep{Roland09}.
\citet{Villata04} note that changes in the Doppler factor can lead to a moderate chromatism if the radiation spectrum is slightly different from the power-law.
Bhatta et al. believed that the local magnetic field enhancement in the jet region, filled electrons with a concave energy spectrum, can lead to the BWB trend in IDV \citep{Bhatta16}.

Based on the available multi-band photometric data, researchers, for example, authors \citep{G-KW92, HuChenGuo14, Poon09, Wu05, Feng20}, conclude that both geometrical effects and physical processes result in IDV of S5~0716+714.
Then, in our opinion, a ``fine tuning'' of parameters is required to explain that these two mechanisms, being independent, produce brightness changes with comparable amplitudes and timescales.

In this paper, we assume that the main part of the blazar S5~0716+714 optical emission comes from the jet region, in which the medium becomes transparent to radiation at the given frequency.
Since the optical thickness of the medium rapidly decreases with distance from the true jet base for high-frequency radiation \citep{BlandfordKonigl79}, then we assume that the radiation at the considered frequencies corresponding to the effective frequencies of the B, V, R$_\text{c}$, I$_\text{c}$ bands comes from the one region.
This region is already optically thin for these frequencies, since the observed spectrum is well described by the power law.
With decreasing frequency, this spectrum flattens and becomes inverted with a spectral index of $5/2$ in its optically thick part \citep{Pachol}.
This region, which we call the optical core by analogy with VLBI jets, is approximately at a constant distance from the true jet base.
Individual parts of the jet flow (main components) pass through this region sequentially.
But since the jet has a helical shape \citep{Lister13,Bach05} and non-radial motion of components \citep{Rast09, Rast11, Rani15}, then for some time the optical core will have an extremely high Doppler factor $\left( \delta\approx30-40\right)$, whereas for a longer time interval $\delta\approx5-10$ \citep{But18a}.
The described change in $\delta$ would be reflected on the long-term light curve of the object, which would have prominent peaks.
There are no such peaks on the observed light curve \citep[e.g., ][]{Raiteri03,Dai13, Dai15, Liao14, Yuan17, XiongBai20}.
It could be explained by the fact that other peaks on the light curve are formed by physical processes. But, there is no enough evidence that variabilities caused by geometrical effects and physical processes are not superimposed.
We explain the absence of prominent peaks, caused by geometrical effects, on the long-term light curve by the fact that individual parts of the main component, which we call sub-components, move with some deviation from the bulk trajectory of the main component.
The Doppler factors of the sub-components $\delta_\chi$ differ from the Doppler factor of the component $\delta$.
On the one hand, this can lead to a flux decrease from the object formed in the component with high $\delta$ and at a small $\delta_\chi$.
On the other hand, the opposite case may occur when the flux from the object increases due to a large $\delta_\chi$ despite a small $\delta$.
One of the indirect confirmations of the presence of small regions with a large Doppler factor is the large brightness temperature, determined from both variability \citep[\(T_\text{br}\sim10^{15}-10^{17}\)~K,][]{WagnerW96, Kraus03, Ostorero06, Fuhrmann08} and observations of the Earth–space radio interferometer \textit{RadioAstron} \citep[\(T_\text{br}>2\cdot 10^{13}\)~K, ][]{Krav20a, Krav20b}.
Another confirmation follows from the \textit{RadioAstron} VLBI map of one of the closest radio galaxies Per~A \citep{Giovannini18}, which jet shows small regions with high brightness.

The continuous formation, development and disappearance of sub-components can explain both the stochastic nature of variability \citep{Amir06, Bhatta13} and several other observational facts.
Namely, the complex and probably composite profiles of flares on both intra- (for example, see Fig.~\ref{fig:lc} for JD~2457130) and inter-night timescales \citep{Bhatta13}.
Different $\delta_\chi$ results in a different offset in the source reference frame of frequencies corresponding to the fixed frequencies in the observer's reference frame.
A convex spectrum of the radiating region leads to the different behavior of the color index in variability caused by an increase in $\delta_\chi$.
For instance, the observed spectrum does not practically change with a small $\delta_\chi$, whereas with an increase in $\delta_\chi$ a stronger BWB trend will appear.
A confirmation of this scenario is the different color index behavior in adjacent IDV events observed by us and in \citep{ZhangWM18}.
The absence of dependence of the BWB trend appearance on the brightness can be explained by a different volume of the radiating region of sub-components.
The different lifetime of sub-components with various volumes and $\delta_\chi$ results in different color index behavior on different time scales.
The sometimes observed micro-oscillations of the S5~0716+714 brightness \citep{Gupta09, Rani10, Bhatta16, Hong18} can be explained by the rotation of the long-lived sub-component around the motion direction of the main component.
Then the different angles between velocity vectors of the sub-component and main component can explain the different oscillation periods.
Modes $n>1$ of the Kelvin-Helmholtz instability \citep{Hardee82} may be responsible for the formation of rotating sub-components.
However, our conclusions are independent of the formation mechanism and the physical nature of sub-components.
As a sub-component we can also imply a certain volume in the blob expanding spherically in the reference frame of the radiating plasma.
Or by a sub-component, we may mean that part of the jet flow whose electrons are accelerated on the slightly curved part of the shock front propagating downstream of the jet.

An alternative assumption is that the observed optical radiation is generated in the optically thin and extended region of the jet.
The fact that bright optical jets can exist was shown by comparing VLBI cores positions with the position of  active galaxies measured by \textit{Gaia} \citep{Kovalev17, PetrovKovalev17, PetrovKovalev19, PlavinKovalev19, KovalevZobnina20}. 
Then the sub-components could also lead to the flux variability, but a change in the color index in variability would only occur with a different spectrum of emitting particles in different main components of the jet.
The latter can be implemented due to the effect of spectral aging \citep{Kardashev62}.
But if we do not impose any additional conditions on the formation of sub-components, their brightness, etc., then the RWB trend in the observational data of S5~0716+714 would be detected more often and for individual events of variability.
It would happen because the RWB trend would appear when the radiation of a sub-component having a steeper power spectrum than the spectrum of the rest radiation dominates.
There are two facts. 
First, the RWB trend for a single variability event was not detected in the S5~0716+714 observations. 
Second, the estimated for IDV and long-term variability frequencies for which the optical depth of the optical core becomes equal to one are approximately consistent with each other. 
These facts support our scenario of the origin of the S5~0716+714 variability on different time scales.

\section{Conclussions}
\label{sec:conclusions}

Our observations occur during the period of strong variability of the blazar S5~0716+714 on a time scale of tens of days.
Analysis of IDV observations for 16 nights out of 23 during 04.2014$-$04.2015 showed the following.
1. The presence of IDV does not depend on either the magnitude or the color index of the object.
2. Most nights with the detected IDV have a BWB trend.
A stronger BWB trend is found in long-term variability.
3. The intra-day correlation coefficient between the magnitude and the color index does not depend on both the magnitude and color index of the object and the variability amplitude.
4. Adjacent IDV events having different color index behavior in variability have been registered.

We assume the observed optical radiation comes from a region where the jet medium becomes transparent to radiation at a given frequency.
This region is approximately at a constant distance from the active nucleus. If the size of this region is comparable to the gravitation radius of a central black hole, having a mass of $5\cdot10^8 M_\odot$, then the transverse component of the magnetic field of this region is $10^2-10^4$~G in the source reference frame.
The frequency of the spectral maximum for the optical emitting region is $\nu_\text{m}=\left(0.6-1.7\right)\cdot 10^{14}$~Hz.
Jet main components moving along the curved paths pass through this region consecutively, producing the long-term variability.
Meanwhile, IDV is formed by sub-components, which are parts of the component moving with some deviation from the general trajectory.
The combination of jet sub-components with a large Doppler factor and the concave synchrotron self-absorption spectrum of the region, from which the observed optical flux comes, simply explains the following facts: (i)  the observed strong variable light curve with no prominent maxima; (ii) the present or absent of the BWB trend, depending on the Doppler factor and the volume of the corresponding sub-component.
Thus, we give a unified explanation for different BWB behavior of the optical variability of the blazar S5~0716+714.
Note that our scenario does not exclude the physical origin of variability but only emphasizes the importance of investigating the influence of geometrical effects on the blazar variability.

\acknowledgments

Author is grateful to Valentina Trofimovna Doroshenko for interest to this remarkable object.
Valentina T. Doroshenko died in July 2017, leaving a great scientific heritage.
She performed all observations analyzed in this article. 

Author is thankful to Elena Nokhrina, Yuri Y. Kovalev, Ilya Pashchenko for the useful discussion of the results of this work.
The theoretical part of this research was supported by the Russian Science Foundation grant 19-72-00105.

\bibliography{Butuzova0716.bib}{}
\bibliographystyle{aasjournal}

\end{document}